\documentclass[fleqn,usenatbib]{mnras}

\usepackage{newtxtext,newtxmath}

\usepackage{subcaption}
\usepackage[T1]{fontenc}
\usepackage{CJK}
\usepackage[whole]{bxcjkjatype} 
\usepackage{CJKutf8}
\usepackage[dvipsnames]{xcolor}
\DeclareRobustCommand{\VAN}[3]{#2}
\let\VANthebibliography\thebibliography
\def\thebibliography{\DeclareRobustCommand{\VAN}[3]{##3}\VANthebibliography}

\usepackage{graphicx}	
\usepackage{amsmath}	

\usepackage{cancel}
\usepackage{soul}

\title[Spectropolarimetry of the type II SN 2021yja]{Spectropolarimetry of the type IIP supernova 2021yja: an unusually high continuum polarization during the photospheric phase
\thanks{Based in part on observations collected at the European Southern Observatory under ESO programs 105.20AU.002 and 108.228K.001.
}
}

\author[S. S. Vasylyev et al.]{
Sergiy S. Vasylyev$^{1,2}\thanks{E-mail: sergiy\_vasylyev@berkeley.edu}$,
Yi Yang\begin{CJK*}{UTF8}{gbsn}
(杨轶)
\end{CJK*}$^{1,3}\thanks{E-mail: yi.yang@berkeley.edu}$,
Kishore C. Patra$^{1,4}$,  
Alexei V. Filippenko$^{1}$,
Dietrich Baade$^{5}$, \newauthor
Thomas G. Brink$^{1,6}$, 
Peter Hoeflich$^{7}$, 
Justyn R. Maund$^{8}$, 
Ferdinando Patat$^{5}$, 
Lifan Wang$^{9}$, \newauthor
J. Craig Wheeler$^{10}$,
WeiKang Zheng$^{1,11}$
\\
$^{1}$Department of Astronomy, University of California, 
Berkeley, CA 94720-3411, USA \\
$^{2}$Steven Nelson Graduate Fellow in Astronomy \\
$^{3}$Bengier-Winslow-Robertson Postdoctoral Fellow in Astronomy \\
$^{4}$Nagaraj-Noll-Otellini Graduate Fellow in Astronomy \\
$^{5}$European Organisation for Astronomical Research in the Southern Hemisphere (ESO), Karl-Schwarzschild-Str.\ 2, 85748 Garching b.\ M{\"u}nchen, Germany \\
$^{6}$Wood Specialist in Astronomy \\
$^{7}$Department of Physics, Florida State University, Tallahassee, Florida 32306-4350, USA \\
$^{8}$Department of Physics and Astronomy, University of Sheffield, Hicks Building, Hounsfield Road, Sheffield S3 7RH, UK \\
$^{9}$George P.\ and Cynthia Woods Mitchell Institute for Fundamental Physics $\&$ Astronomy, Texas A$\&$M University, 4242 TAMU, College Station, TX 77843, USA \\
$^{10}$Department of Astronomy, University of Texas, Austin, TX 78712, USA \\
$^{11}$Eustace Specialist in Astronomy \\
}

\date{Accepted XXX. Received YYY; in original form ZZZ}

\pubyear{2023}

\begin{document}
\label{firstpage}
\pagerange{\pageref{firstpage}--\pageref{lastpage}}
\maketitle

\begin{abstract}
We present six epochs of optical spectropolarimetry of the Type IIP supernova (SN) 2021yja ranging from $\sim 25$ to 95 days after the explosion. 
An unusually high continuum linear polarization of $p \approx 0.9\%$ is measured during the early photospheric phase, followed by a steady decrease well before the onset of the nebular phase. This behaviour has not been observed before in Type IIP supernovae (SNe~IIP).
The observed continuum polarization angle does not change significantly during the photospheric phase. We find a pronounced axis of symmetry in the global ejecta that is shared in common with the H$\alpha$ and Ca~II near-infrared triplet lines. These observations are consistent with an ellipsoidal geometry. The temporal evolution of the continuum polarization is also compatible with the SN ejecta interacting with aspherical circumstellar matter, although no spectroscopic features that may be associated with strong interaction can be identified. 
Alternatively, we consider the source of the high polarization to be an extended hydrogen envelope that is indistinguishable from low-density circumstellar matter. 
\end{abstract}





\section{Introduction}
\label{s:intro}

It is widely accepted that the core collapse of a star with zero-age main-sequence (ZAMS) mass $\geq 8\,M_{\odot}$ produces hydrogen-rich Type II and a subset of hydrogen-poor Type I (Ib/Ic; stripped-envelope) supernovae (SNe). Although most of these stars explode as SNe, some models predict direct collapse to a black hole in certain progenitor mass ranges \citep{sukhbold_core-collapse_2016,byrne_nothing_2022}. SNe~II are also distinguished by their light-curve shape; 
an SN~II with a linearly (in magnitudes) declining light curve is designated as IIL, whereas an SN~II with an extended plateau lasting $\sim 90$\,days after explosion is classified as IIP; see \citet{filippenko_optical_1997} and \citet{gal-yam_observational_2017} for reviews of SN classification.
However, the  distinction between SNe~IIP and IIL is not clear, with recent works suggesting that these subtypes instead constitute a continuum \citep{anderson_characterizing_2014,valenti_diversity_2016}.

The process that drives the explosions of core-collapse supernovae (CCSNe) also remains unclear. 
One of the most promising models is a neutrino-driven mechanism  \citep{bethe_revival_1985,janka_explosion_2012}. 
However, numerical simulations based on this mechanism fail to reproduce the energies observed in CCSNe  \citep{melson_neutrino-driven_2015}. 
The discrepancy between the simulated and the observed energies in CCSNe could be reconciled by an enhancement of the neutrino heating efficiency caused by multidimensional hydrodynamic instabilities. 
Possible sources include convective motion, standing-accretion-shock instability (SASI; \citealp{blondin_stability_2003, marek_delayed_2009}), and Rayleigh-Taylor instability \citep{kifonidis_non-spherical_2003}. In the case of SNe~II, such instabilities may introduce inhomogeneities in the core-collapse process and at the interface between the inner ejecta and the hydrogen envelope. 
Additional instabilities may be induced by effects of general relativity, rotation, and magnetohydrodynamics near and around the core \citep{leblanc_numerical_1970}. On the other hand, a jet-driven bipolar explosion model has also been invoked to explain the observed energies and asphericities in SNe~II \citep{khokhlov_jet-induced_1999,wang_axisymmetric_2002,chugai_asymmetry_2006,maund_spectropolarimetry_2007,couch_aspherical_2009,papish_call_2015,mauerhan_asphericity_2017}. 
It is conceivable that the explosion process may involve a combination of these two models. In order to test these hypotheses, methods have been developed to study the internal structure of SNe and characterise the inhomogeneities of their ejecta. 

Spectropolarimetry, which involves counting polarized photons at
different wavelengths, is an extremely useful probe of the geometric
properties of supernova (SN) ejecta and the degree of chemical asphericity
without spatially resolving the source. Furthermore, polarization
measured across prominent spectral lines can characterise the
distribution of individual elements within the ejecta (see Section \ref{s:qu_diagrams} for discussion). 
Reviews of polarimetric studies of SNe include \citet{wang_spectropolarimetry_2008}, \citet{branch_supernova_2017}, and \citet{patat_introduction_2017}. 
The general principle behind this method is as follows. Polarized photons are produced via Thomson scattering in SN atmospheres. The electric field of the scattered photon is perpendicular to the plane of scattering. 
Any deviation from spherical symmetry would result in an incomplete cancellation of the electric-field vectors (``E-vectors"). Since the SN is observed as a point source and the light hitting a detector is an integration of all the E-vectors, this asphericity would result in a net polarization that is always greater than zero.

Early studies of SNe~IIP showed characteristically low polarization during the photospheric phase \citep{leonard_is_2001,leonard_non-spherical_2006,chornock_large_2010}.
During the photospheric (or the light-curve plateau) phase, the outer H-rich envelope, which exhibits a high electron-scattering opacity, reemits the thermal energy deposited by the SN shock through a recombination process. Therefore, polarimetry of SNe~IIP during the plateau phase measures the geometry of the electron-scattering atmosphere as it recedes inward through the H-rich envelope. Asymmetries in the inner core are not visible at this phase owing to the high electron-scattering opacity in the envelope. 

As the SN expands and cools adiabatically, the photosphere recedes deeper into the stratified ejecta. This allows one to peer into the different layers of the explosion as time progresses. The observed level of polarization as a function of wavelength can be used to infer the degree of inhomogeneities and/or asymmetries in the ejecta (see, e.g., \citealt{tanaka_three-dimensional_2012} and \citealt{tanaka_three-dimensional_2017}). 
In particular, a wavelength-independent continuum polarization is most likely to be produced by Thomson scattering of free electrons that deviate from a spherically symmetric distribution. It may also be caused by the presence of energy sources that are offset from the kinematic centre of the explosion
\citep{chugai_polarization_1992,Hoeflich_etal_1995, kasen_analysis_2003, Hoeflich_etal_2006}. Polarization signals across spectral lines can be understood by unevenly blocking of the photosphere by the associated line opacity, and by its frequency variations in the thermalisation depth \citep{Hoeflich_Yang_2023}, tracing the distribution of elements in the SN ejecta.
Therefore, time-series spectropolarimetry can tomographically measure the degree of asphericity from the outer to inner layers of the SN ejecta. The geometry of the inner core is dependent on the degree of instabilities in the SN explosion, further placing constraints on the explosion mechanisms discussed above. 

Previous spectropolarimetric studies of SNe~IIP such as SNe
2004dj, 2006my, 2006ov, and 2007aa show evidence for the emergence of an aspherical core following the photospheric phase \citep{leonard_non-spherical_2006,chornock_large_2010}. 
In contrast to the generally seen low continuum polarization intrinsic to SNe~IIP,  a few notable exceptions include the Type IIP/L SN\,2013ej \citep{mauerhan_asphericity_2017,nagao_evidence_2021}, which displayed a significant continuum polarization at the earliest epochs, and SN\,2017gmr \citep{nagao_aspherical_2019}, which exhibited a steep rise in polarization as early as 30 days before the plateau dropoff. 
Another peculiar case was observed for SN\,2017ahn \citep{nagao_evidence_2021}, where the low continuum polarization during the photospheric phase persisted well into the nebular phase. The sample of SNe~IIP for which there have been spectropolarimetric measurements remains small, despite being the most commonly observed type of SNe. This work is part of an effort to obtain high-quality data for nearby CCSNe  with low interstellar polarization (ISP) contamination. 

SN\,2021yja was discovered on 8 Sep. 2021 at 13:12:00 (UTC dates are used throughout this paper) in the spiral galaxy NGC\,1325 \citep{smith_atlas21bidw_2021} by the Asteroid Terrestrial-impact Last Alert System \citep[ATLAS;][]{tonry_atlas_2018}. 
A redshift of $z = 0.005307$ reported by \citet{Springob_etal_2005} was used to correct for the redshift of the host galaxy. 
We adopt the time of explosion 
($t_{\text{exp}}$) to be 7.5 Sep. 
2021, as estimated by 
\cite{vasylyev_early-time_2022}, 
which is roughly the midpoint 
between the last nondetection on 
6.48 Sep. 2021 and the first 
detection on 8.55 Sep. 2021.
All phases will be given in days relative to this date throughout the paper. 
Previous photometric and spectroscopic studies of SN\,2021yja have found weak mass loss from the progenitor star and little evidence of the presence of dense circumstellar matter \citep[CSM;][]{hosseinzadeh_weak_2022, vasylyev_early-time_2022}.

SN\,2021yja was observed to have a relatively long plateau phase ($\sim 140$ days), which can be explained by a hydrogen-envelope mass of $\geq 5\,M_{\odot}$ \citep{hosseinzadeh_weak_2022}.
However, there are some signatures of CSM interaction, including X-ray and radio emission \citep{alsaberi_radio_2021,ryder_radio_2021}, an early-time optical flux excess, and a short rise time (see \citealp{kozyreva_circumstellar_2022}),
but we do not expect a significant contribution to the polarization signal through free electron scattering in the low-density CSM. 
We note that \citet{kozyreva_circumstellar_2022} suggest that the progenitor of SN\,2021yja likely possessed a convective envelope. 
Following a radiative transfer hydrodynamical modeling procedure as discussed by \citet{kozyreva_role_2019}, this convective red supergiant envelope may build up a pre-existing CSM of $0.55~M_{\odot}$ at a radius $r \approx 2 \times 10^{14}$\,cm (or $\sim 2700\,R_{\odot}$).
The progenitor star's initial mass was constrained to be $\sim 15\,M_{\odot}$ and have a pre-explosion radius of $\sim 630\,R_{\odot}$ \citep{kozyreva_circumstellar_2022}.
The $^{56}$Ni mass produced in the explosion was estimated to be $\sim 0.12\,M_{\odot}$ by \citet{hosseinzadeh_weak_2022} and \citet{vasylyev_early-time_2022}, which is more than 1$\sigma$ above the typical value $\sim 0.044\,M_{\odot}$ for SNe~II \citep{anderson_meta-analysis_2019}. 
The explosion energy was determined to be (1--3) $\times 10^{51}$\,ergs using photospheric-velocity values from radiative-transfer modeling \citep{vasylyev_early-time_2022}.  We note a discrepancy between the velocity measurements in the aforementioned work and that of \citet{hosseinzadeh_weak_2022}, who determine a lower value using the blueshift of the \ion{Fe}{II} absorption minimum.

This paper is organised as follows. In Section \ref{s:meth}, we present the summary of our spectropolarimetric observations and discuss systematic errors associated with instrumentation and contamination from interstellar polarization. We calculate the continuum linear polarization and polarization angle in Section \ref{s:specpol}, discuss the uniqueness of SN\,2021yja among the population of SNe~II with spectropolarimetric observations, and interpret the geometry of the SN ejecta. A brief summary of the study is given in Section~\ref{sec:conclusions}.
With this work, we aim to signal a paradigm shift from considering SNe~II as a homogeneous population to instead recognising the wide range of possible geometries needed to explain the emerging diversity in spectropolarimetric observations. 


\section{Summary of Observations and Data Analysis}\label{s:meth}

\subsection{VLT Spectropolarimetry}
We obtained one epoch of spectropolarimetry of SN\,2021yja using the FOcal Reducer and low-dispersion Spectrograph 2 (FORS2; \citealp{appenzeller_successful_1998}) on the Unit Telescope~1 (UT1, Antu) of the ESO Very Large Telescope (VLT). 
Observations were made in the Polarimetric Multi-Object Spectroscopy (PMOS) mode on 02 Oct. 2021, corresponding to $\sim 25$ days after the estimated time of the explosion. 
Grism 300V and a 1$\arcsec$-wide slit were adopted, resulting in a spectral resolving power of $R \approx 440$, corresponding to the size of a resolution element of 13\,\AA\ at a central wavelength of 5849~\AA\ (see, e.g., \citealp{anderson_etal_2018}). 
The slit was aligned with the north celestial meridian considering all observations were carried out at small airmass ($\lesssim$1.2, see Table~\ref{tbl:specpol_log}). Given the presence of a linear atmospheric dispersion compensator (LADC; \citealp{Avila_etal_1997}), we consider the small misalignment between the slit and the parallactic angle \citep{filippenko82} to have a negligible effect on the observed spectral energy distribution.

The VLT spectropolarimetric observation consisted of four exposures each with the half-wave retarder plate positioned at angles of $0^{\circ}$, $45^{\circ}$, $22.5^{\circ}$, and $67.5^{\circ}$. 
The total 900\,s $\times$ 4 retarder-plate angles integration was split into three sets of 300\,s exposures to reduce the impact of cosmic rays.
After bias subtraction and flat-field corrections, the ordinary (o) and extraordinary (e) beams were extracted following standard routines within IRAF\footnote{IRAF is distributed by the National Optical Astronomy Observatories, which are operated by the Association of Universities for Research in Astronomy, Inc., under cooperative agreement with the U.S. National Science Foundation (NSF).}. 
Wavelength calibration was done separately for the o-ray and e-ray in each individual frame with root-mean-square (RMS) accuracy better than $\sim 0.2$\,\AA.
Each beam has been flux-calibrated using the flux standard star EGGR 141 observed at an airmass of 1.0  the following night, with the identical polarimetry optics used to observe SN\,2021yja in place and at the half-wave plate angle $0^\circ$.

Stokes parameters, bias-corrected polarization degrees, and associated uncertainties were computed with our own routines following the procedures of \citet{patat_error_2006} and \cite{maund_spectropolarimetry_2007}. 
The instrumental polarization of FORS2/PMOS ($\lesssim 0.1$\%) was further corrected following \citet{cikota_linear_2017}. 
Details of the reduction of FORS spectropolarimetry can be found in the FORS2 Spectropolarimetry Cookbook and Reflex Tutorial\footnote{\url{ftp://ftp.eso.org/pub/dfs/pipelines/instruments/fors/fors-pmos-reflex-tutorial-1.3.pdf}}, \citet{cikota_linear_2017}, and in Appendix~A of \citet{yang_young_2020}. A test of the precision and stability of the FORS2 instrument for spectropolarimetry is conducted in Appendix~\ref{section_sanity} of this paper.

\subsection{Kast Spectropolarimetry and Lick Photometry}
\label{s:kast_kait} 
Spectropolarimetry of SN\,2021yja was obtained using the polarimetry mode of the Kast Double Spectrograph on the Shane 3\,m telescope at Lick Observatory \citep{miller_ccd_1988,miller_stone_1994}. 
The Kast spectropolarimetry sequence of SN\,2021yja contains five epochs of observations that span days 30 to 95 after the explosion.
Observations and data reduction were carried out following the
description provided by \citet{patra_spectropolarimetry_2021}. 
The 300 lines\,mm$^{-1}$ grating and the 3$\arcsec$-wide slit were adopted, resulting in a spectral resolving power of 
$R \approx 380$, corresponding to the size of a resolution element 
of 18\,\AA\, at a central wavelength of 6800~\AA.
A log of the VLT and Kast spectropolarimetry of SN\,2021yja is presented in Table~\ref{tbl:specpol_log}. 

In order to characterise the instrumental polarization, nightly observations of the low-polarization standard star HD\,212311 were carried out with the same polarimetry optics used to observe SN\,2021yja. 
The average normalised Stokes $q$ and $u$ values\footnote{Following the definition adopted by \citet{patra_spectropolarimetry_2022}, where $q = Q/I$ and $u = U/I$.} over all epochs were measured to be $<0.05\%$, 
thus confirming negligible instrumental polarization and long-term stability of the Kast spectropolarimeter.

\begin{table*}
\caption{Journal of spectropolarimetric observations of SN\,2021yja.} 
\begin{tabular}{cccccccc}
	\hline 
	\hline
	UT Date & MJD$^b$ & Phase$^a$ & Instrument & Airmass Range & Avg. Seeing &  Exposure Time$^c$   \\ 
	(MM-DD-YYYY)&   & (days) & &  & (arcsec) & (s)   \\ 
	\hline 
    10-02-2021 & 59489.20 & 25 & VLT/FORS2 & 1.24--1.05 & 1.06--0.74 & $300 \times 4 \times 3$ \\
    10-07-2021 &59494.46& 30 & Kast & 2.0--2.2 & 2 & $600 \times 4 \times 1$ \\
    10-15-2021 &59502.48& 38 & Kast & 2.0--2.1 & 2 &  $600 \times 4 \times 3$\\
    11-03-2021 &59521.45& 57 & Kast & 1.98--2.07 & 1.2 &  $600 \times 4 \times 3$ \\
    11-12-2021 &59530.28& 66 & Kast & 1.98--2.07 & 1.2 &  $600 \times 4 \times 3$ \\
    12-11-2021 &59559.24& 95 & Kast & 1.9--2.2 & 1.3 &  $600 \times 4 \times 3$\\
    
	\hline 
\end{tabular}\\
{$^a$}{Days after the estimated time of explosion on MJD 59464 (7 Sep. 2021).} \\
{$^b$}{MJD is given as the start time of the CCD exposure.}
{$^c$}{Exposure time of a single exposure $\times$ 4 retarder-plate angles $\times$ number of loops. Wavelength ranges for Kast and VLT/FORS2 are 4570--9800\,\AA, and 3500--9100\,\AA, respectively.} 
\label{tbl:specpol_log}
\end{table*} 

Follow-up photometry of SN\,2021yja was performed by the 0.76\,m Katzman Automatic Imaging Telescope (KAIT) as part of the Lick Observatory Supernova Search \citep[LOSS;][]{filippenko_loss_2001}, as well as the 1\,m Nickel telescope at Lick Observatory \citep{mauerhan_asphericity_2017}. 
$\textit{BVRI}$-band light curves from these observations are presented by \citet{vasylyev_early-time_2022}.

\subsection{Interstellar Polarization}
\label{sec:isp}
The polarization that is intrinsic to the explosion of SN\,2021yja could be contaminated by the polarization caused by interstellar aspherical dust grains along the SN-Earth line of sight. In particular, such an interstellar polarization (ISP) can be induced by the dichroic extinction owing to nonspherical paramagnetic dust grains which are partially aligned by a large-scale magnetic field. 
Therefore, the level of such ISP needs to be estimated in order to isolate the intrinsic polarization of SN\,2021yja. 
In the case of this study, several observational properties suggest a low level of interstellar polarization toward SN\,2021yja.

First, the reddening caused by Galactic dust toward the SN\,2021yja line of sight can be estimated as $E(B-V) = 0.02$\,mag using the NASA/IPAC NED Galactic Extinction Calculator based on the $R_{V} = 3.1$ reddening law and the extinction map from \citet{schlafly_measuring_2011}. 
According to the empirical upper limit of polarization induced by the 
dichroic extinction of interstellar dust grains, $p_{\rm ISP} \textless 9 \times E(B-V)$\%, we estimate  the ISP contributed by the Galactic dust to be $p_{\rm MW}^{\rm ISP}\lesssim 0.18$\% \citep{serkowski_wavelength_1975}. 
Although the ISP within the host galaxy of SN\,2021yja remains unclear, we can still estimate the total line-of-sight ISP from the emission component of the strong P~Cygni profile of the Balmer lines. The central region of the H$\alpha$ emission is intrinsically unpolarized in the absence of a strong magnetic field since the emitted photons are dominated by recombination and resonance scattering rather than Thomson scattering \citep{mccall_are_1984,wang_hydrogen_2004}. The minimum level of polarization across the H$\alpha$ emission reaches 0.046\%\,$\pm$\,0.042\% and 0.047\%\,$\pm$\,0.039\% as measured from 25\,\AA\ and 30\,\AA\ binned VLT spectra at day 25, respectively, which are all roughly consistent with zero within their 1$\sigma$ uncertainties. 
Therefore, we assume that the interstellar polarization toward SN\,2021yja is negligible.


\begin{figure*}
    \centering
    \includegraphics[width=0.9\textwidth]{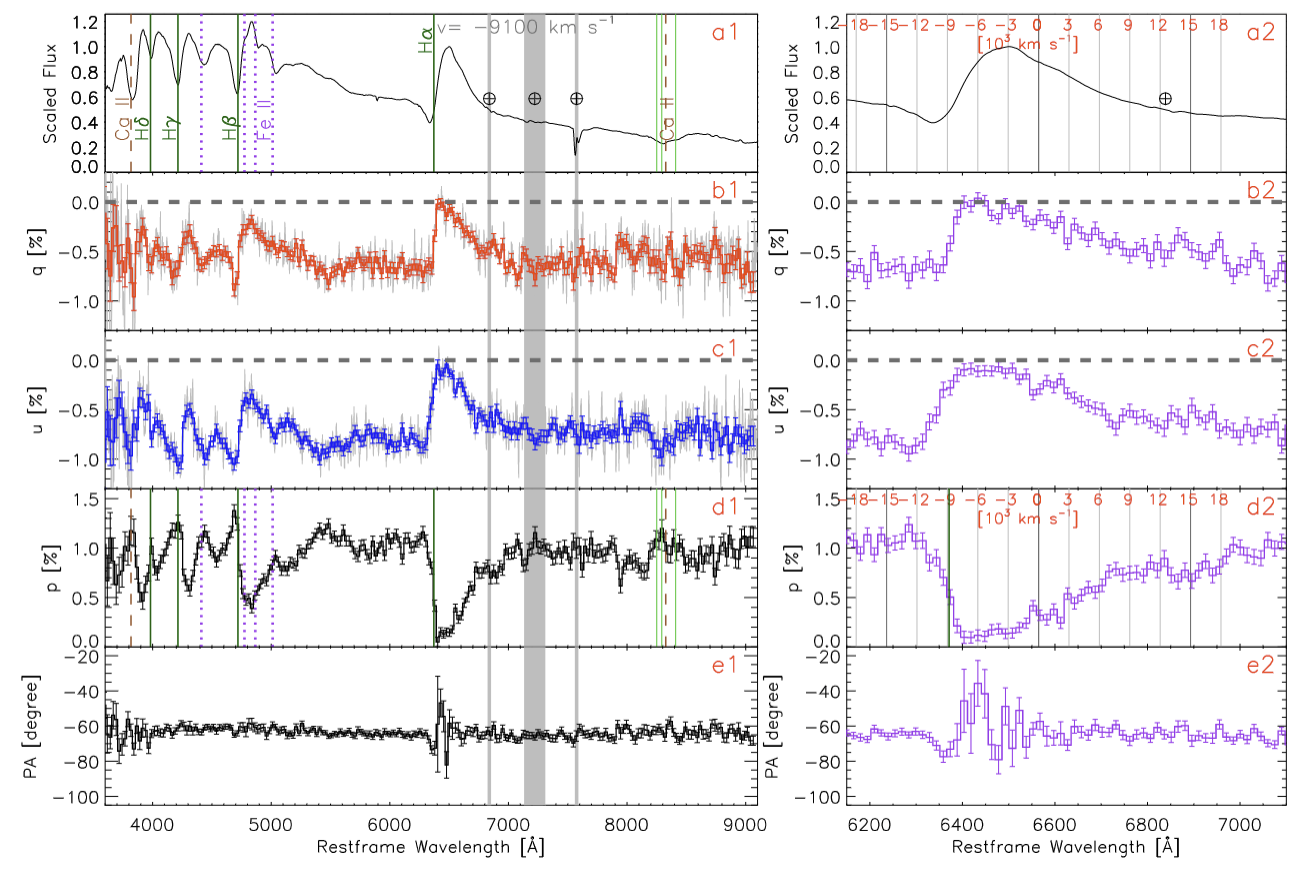}
\caption{{{\it Left column:} spectropolarimetry of SN\,2021yja obtained on day +25 (epoch 1) using VLT FORS2. The five subpanels (from top to bottom) show (a1) an arbitrarily-scaled total-flux spectrum with major spectral lines identified; (b1, c1) the normalised Stokes $q$ and $u$, respectively; (d1) the polarization spectrum ($p$); and (e1) the polarization position angle (PA). 
The data in the left panel have been rebinned to 25\,\AA\ for better illustration. Some major telluric lines are labeled by $\earth$ and vertical grey-shaded bands. {\it Right column:} similar to the left column, but zoomed in on the H$\alpha$ profile. The polarization data have been rebinned to 15\,\AA\ for the purpose of presentation. Vertical grey lines together with the red tick marks indicate velocities relative to H$\alpha$ in the rest frame.}
}
    \label{fig:qu_vlt_yi}
\end{figure*}

\begin{figure*}
   \begin{minipage}[t]{0.49\textwidth}
     \centering
     \includegraphics[width=1.12\linewidth]{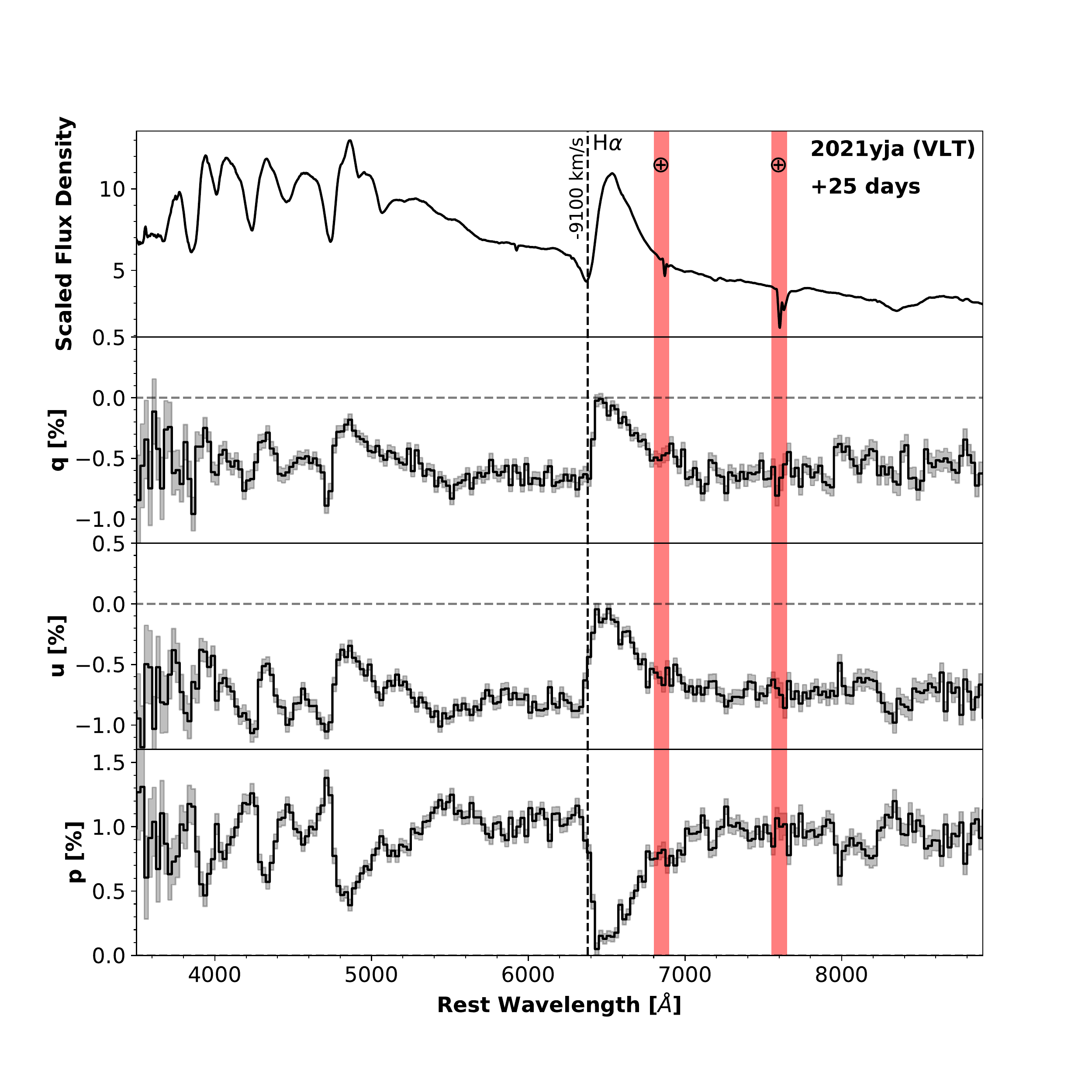}
     \caption{VLT spectropolarimetry of SN\,2021yja on 02 Oct. 2021 (day $+$25). The four panels (from top to bottom) present the scaled flux-density spectrum; the normalised, ISP-corrected Stokes parameters $q$ and $u$, and the polarization spectrum $p$, respectively. The polarization data have been rebinned to 25\,\AA\ for clarity.  
     The vertical size of each bin of the grey-shaded histograms indicates the $1\sigma$ uncertainty. 
     The vertical blue-dashed line marks the blueshifted absorption minimum of the P~Cygni profile of H$\alpha$ at an expansion velocity of 9100\,km\,s$^{-1}$. Some major telluric lines are labeled by red-shaded lines marked with $\earth$.}\label{fig:pol_ep1}
   \end{minipage}\hfill
   \begin{minipage}[t]{0.49\textwidth}
     \centering
     \includegraphics[width=1.12\linewidth]{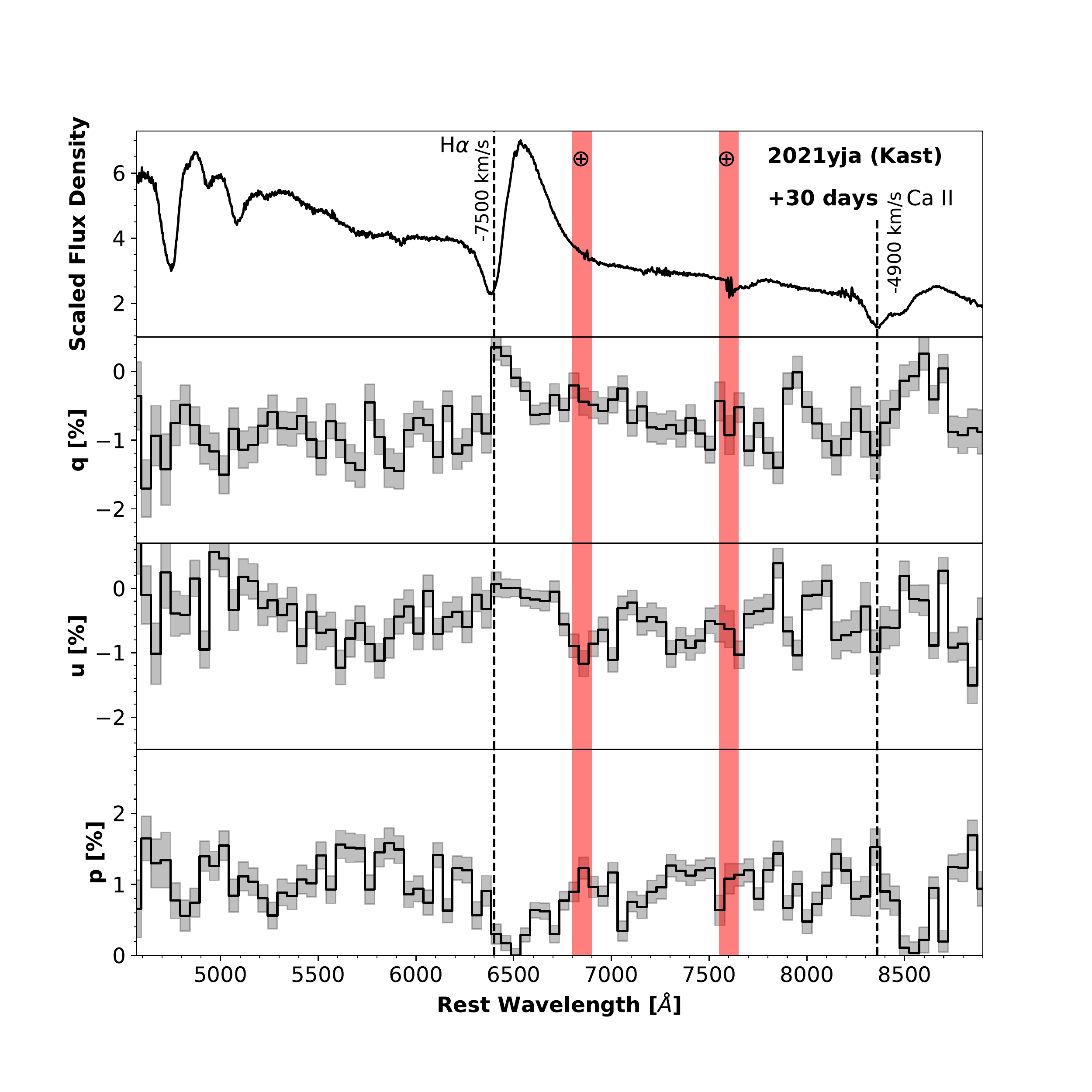}
     \caption{Kast spectropolarimetry of SN\,2021yja on 07 Oct. 2021 (+30 days). Major telluric lines are marked by vertical red columns. Polarization data have been binned to 50\,\AA. }\label{fig:pol_ep2}
   \end{minipage}
   \begin{minipage}[t]{0.49\textwidth}
     \centering
     \includegraphics[width=1.12\linewidth]{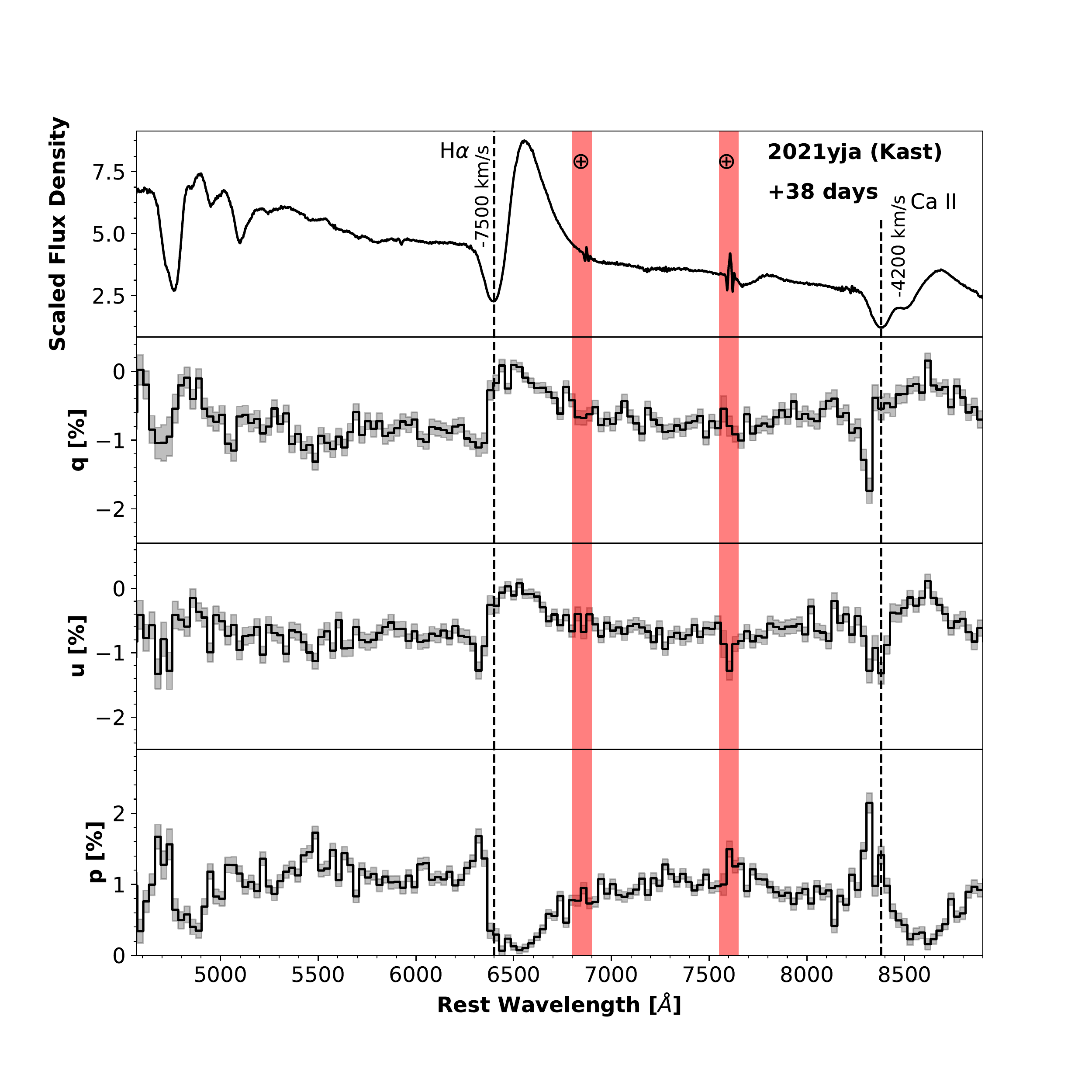}
     \caption{Same as Figure~\ref{fig:pol_ep2} but for 15 Oct. 2021 (+38 days)}\label{fig:pol_ep3}
   \end{minipage}
   \begin{minipage}[t]{0.49\textwidth}
     \centering
     \includegraphics[width=1.12\linewidth]{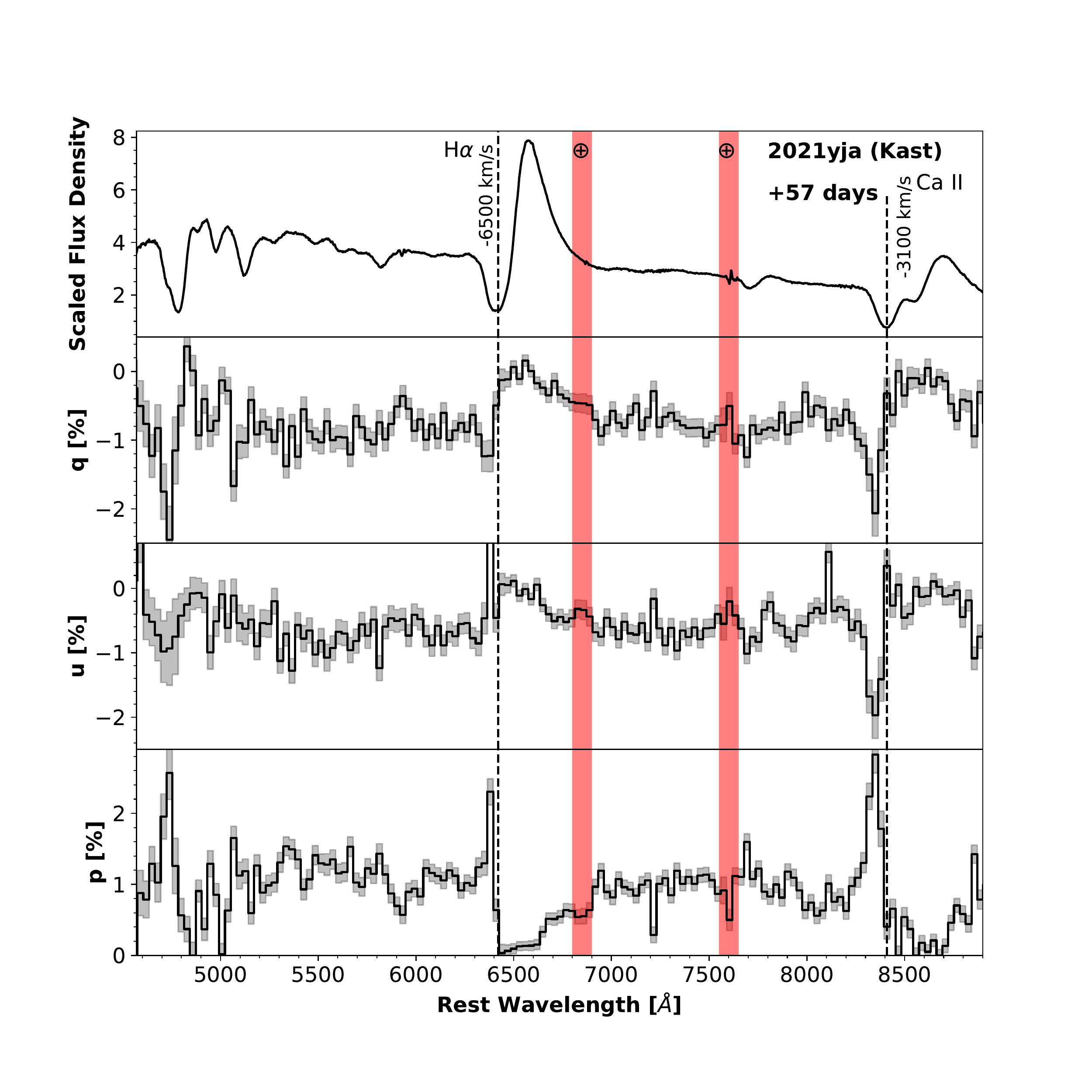}
     \caption{Same as Figure~\ref{fig:pol_ep2} but for 03 Nov. 2021 (+57 days).}\label{fig:pol_ep4}
   \end{minipage}
\end{figure*}

\begin{figure*}
   \begin{minipage}[t]{0.49\textwidth}
     \centering
     \includegraphics[width=1.12\linewidth]{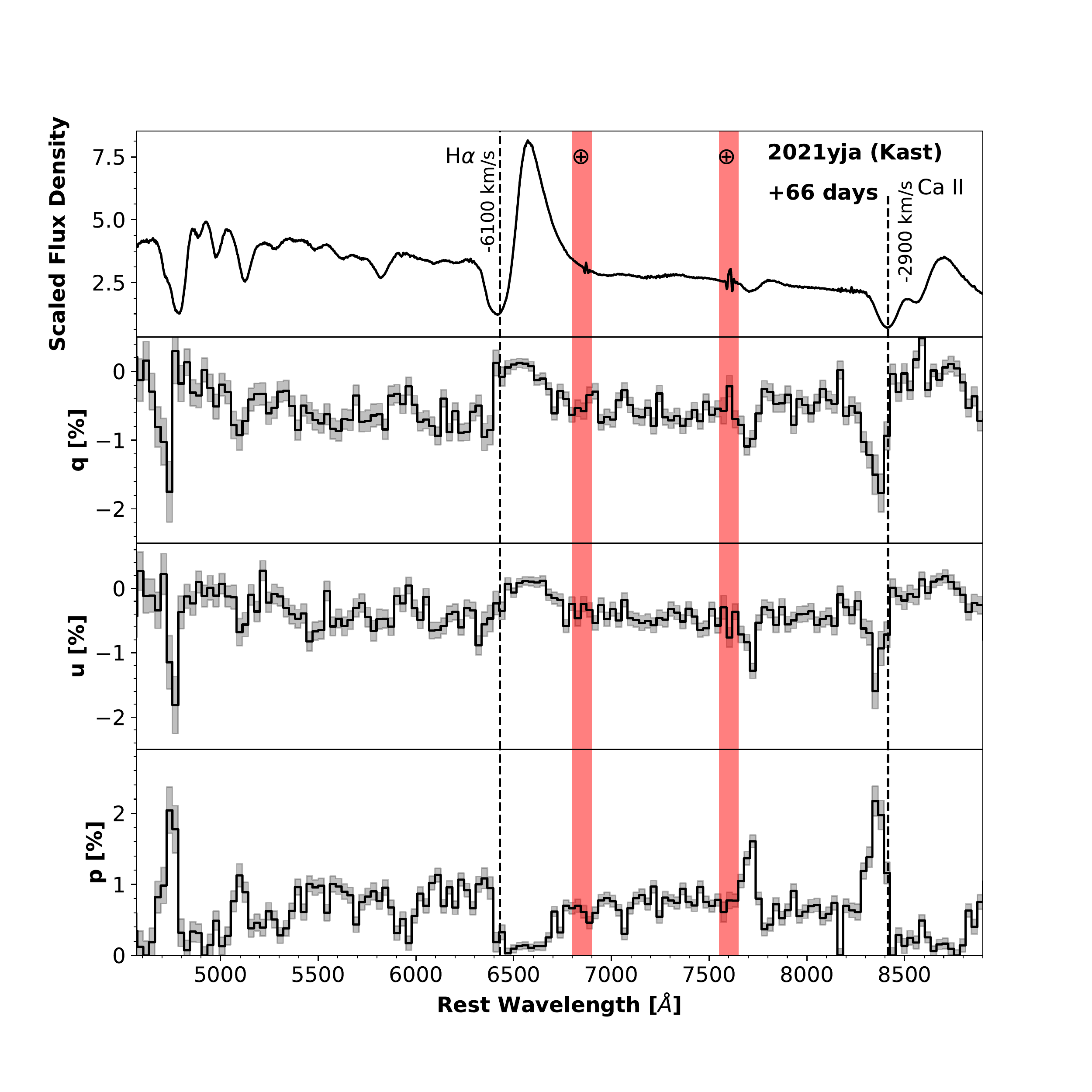}
     \caption{Same as Figure~\ref{fig:pol_ep2} but for 12 Nov. 2021 (+66 days).}\label{fig:pol_ep5}
   \end{minipage}\hfill
   \begin{minipage}[t]{0.49\textwidth}
     \centering
     \includegraphics[width=1.12\linewidth]{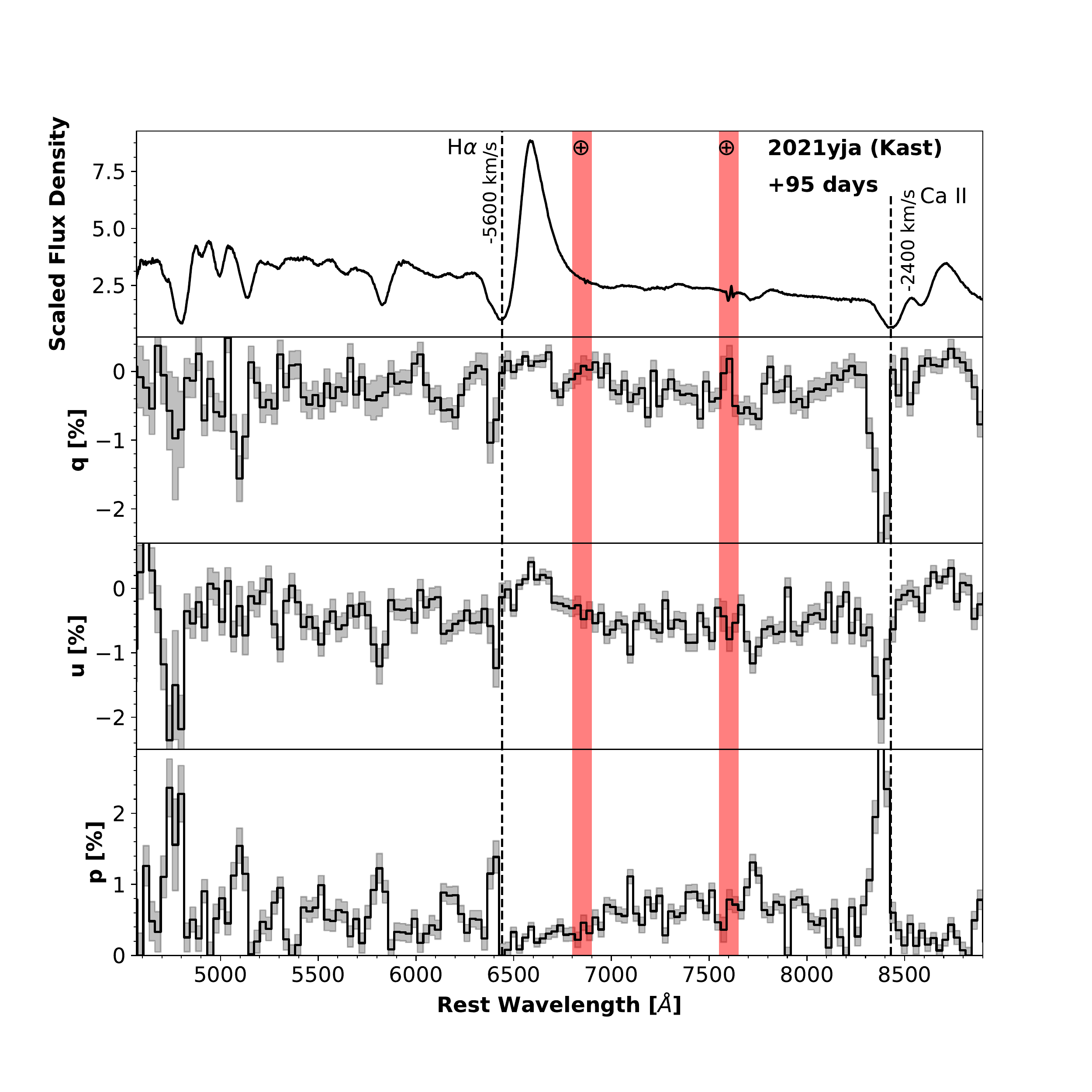}
     \caption{Same as Figure~\ref{fig:pol_ep2} but for 11 Dec. 2021 (+95 days).}\label{fig:pol_ep6}
   \end{minipage}
\end{figure*}

\begin{figure*}
    \includegraphics[width=0.9\textwidth]{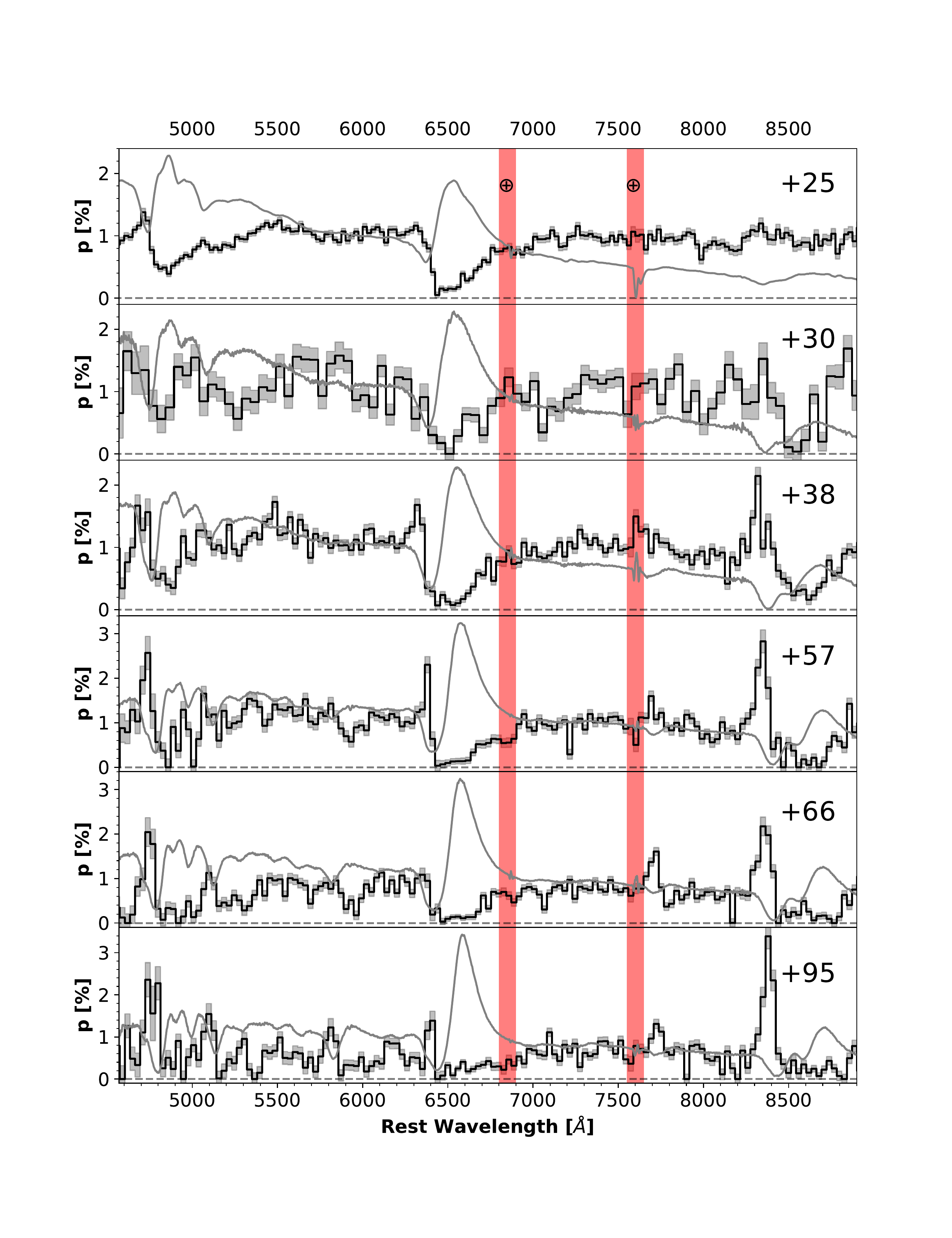}
    \caption{
    Summary of VLT (rebinned to 25\,\AA) and Kast spectropolarimetry (rebinned to 50\,\AA\ for day +30  and 30\,\AA\ for days +38 to +95) of SN~2021yja. The arbitrarily-scaled total-flux spectrum is plotted in grey for each of the observations. Major telluric lines are indicated by \earth. }
    \label{fig:all_epochs_pol_rows}
\end{figure*}

\begin{figure}
    \centering
\includegraphics[width=0.53\textwidth]{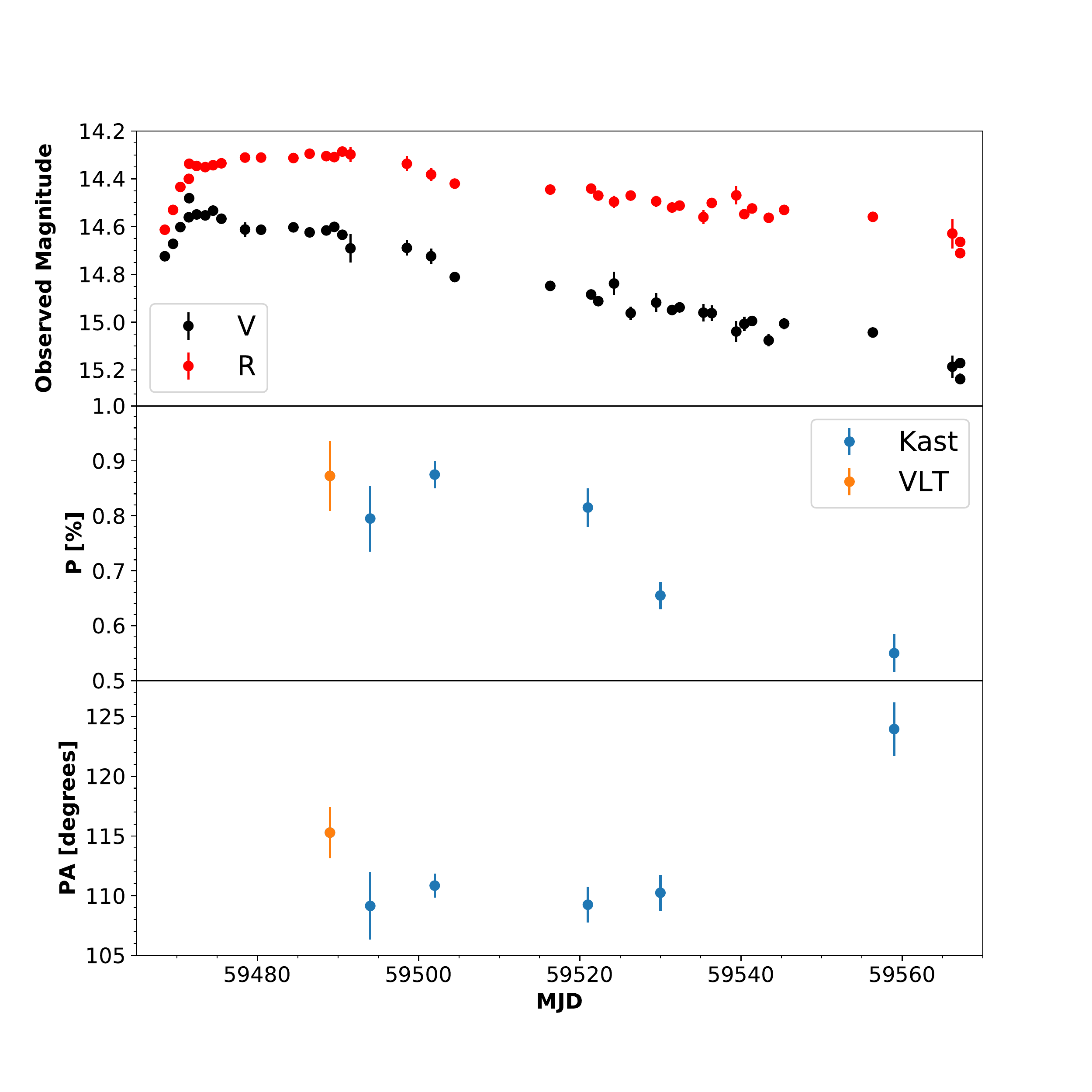}
    \caption{VLT/Kast polarimetry and KAIT $VR$-band photometry. The continuum polarization is calculated across 6800--7200\,\AA\ and 7820--8140\,\AA\ in order to exclude prominent telluric features. Error bars represent 1$\sigma$ uncertainties.
    }
    \label{fig:cont}
\end{figure}

\begin{table}
\caption{Continuum polarization and the associated polarization position angle estimated for SN\,2021yja.}
\begin{tabular}{ccccccc}
	\hline 
	\hline
	Phase$^a$ & Telescope / & Pol$^b$ & Pol. error & PA & PA err   \\ 
	(days) & Instrument & (\%) & ($\sigma$\,\%) & (deg) & ($\sigma$\,deg)   \\ 
	\hline 
    25 & VLT / FORS2 & 0.87 & 0.06 & 115.3 & 2.1 \\
    30 & Shane / Kast & 0.80 & 0.06 & 109.2 & 2.8 \\
    38 & Shane / Kast & 0.88 & 0.03 & 110.9 & 1.0 \\
    57 & Shane / Kast & 0.82 & 0.04 & 109.3 & 1.5 \\
    66 & Shane / Kast & 0.66 & 0.03 & 110.3 & 1.5 \\
    95 & Shane / Kast & 0.55 & 0.04 & 124.0 & 2.3 \\
	\hline 
\end{tabular}\\
{$^a$}{Phases are given in days after the explosion.}.\\
{$^b$}{Continuum polarization and PA were computed based on the $1\sigma$-error-weighted Stokes parameters with wavelength ranges 6800--7200\,\AA\ and 7820--8140\,\AA.}\\
\label{tbl:contpol_log}
\end{table} 

\begin{figure*}
    \centering
    \includegraphics[width=0.9\textwidth]{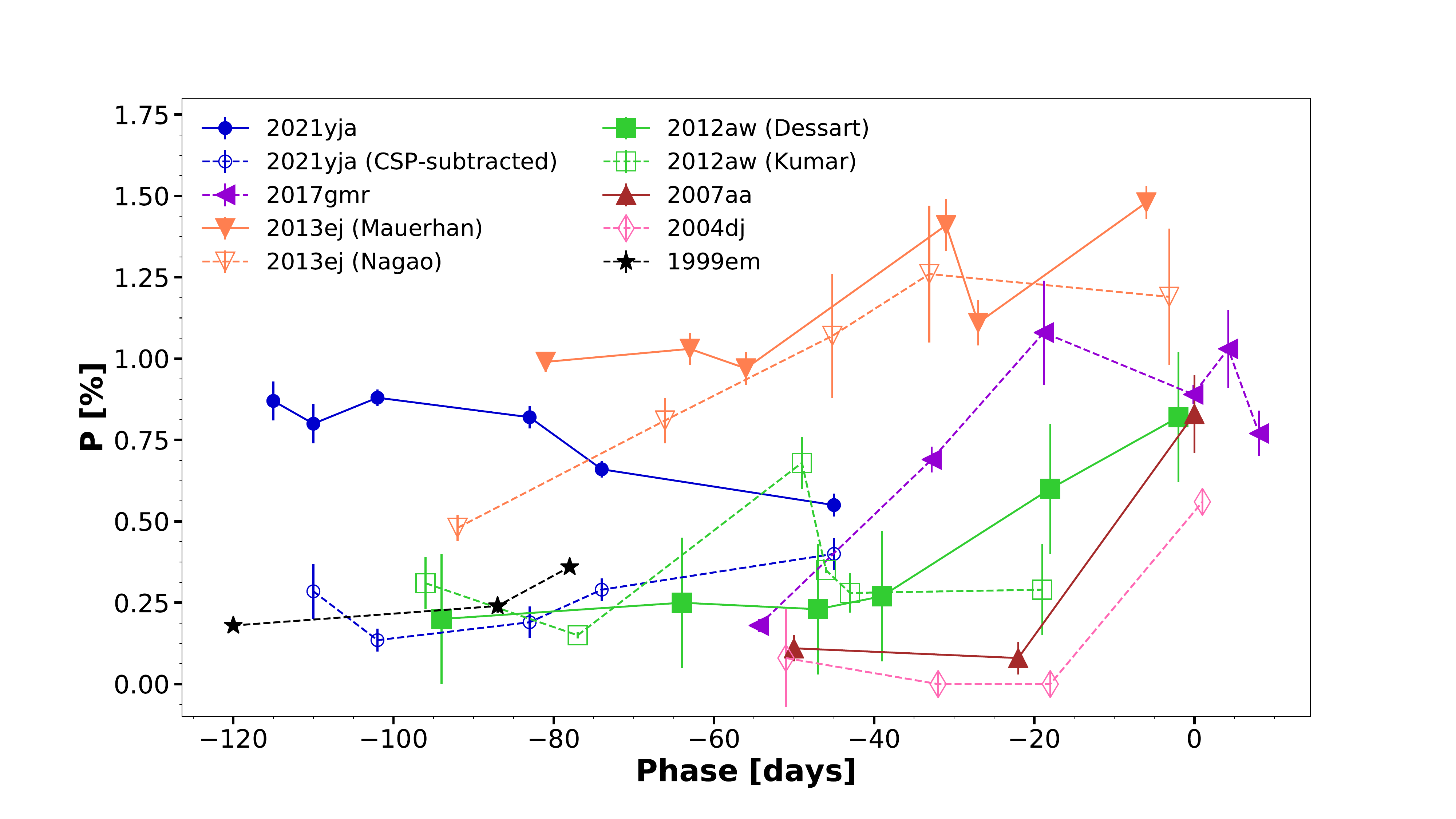}
    \caption{Continuum polarization of SN\,2021yja compared with that of Type II SNe 1999em (\citealp{leonard_is_2001}; black star), 2004dj (\citealp{leonard_non-spherical_2006}; pink diamond), 2007aa (\citealp{chornock_large_2010}; brown triangle), 2012aw (\citealp{dessart_multiepoch_2021,kumar_broad-band_2014}; green square), 2013ej (\citealp{mauerhan_asphericity_2017, nagao_evidence_2021}; orange triangle), and 2017gmr (\citealp{nagao_aspherical_2019}; purple triangle). 
    For each SN, the phase is relative to the end of its plateau, which is defined as the midpoint of the rapidly declining light-curve dropoff. Error bars represent 1$\sigma$ uncertainties. 
    }
    \label{fig:sne_comp}
\end{figure*}

\section{Spectropolarimetry}
\label{s:specpol}
In the following subsections, we present the VLT and Kast spectropolarimetry of SN\,2021yja obtained during six epochs that span days $+25$ to $+95$. First, we focus on the peculiar, high level of continuum polarization at early photospheric phases. Second, we compare the intrinsic polarization properties of SN\,2021yja with a sample of Type IIP SNe that have multi-epoch polarimetry measurements. Next, we discuss the effects on the observed polarization by an additional heating source. Finally, we conduct a detailed investigation of the polarization profile across the H$\alpha$ line and a possible late-time rise of the \ion{Ca}{ii} near-infrared triplet (air wavelengths 8498.02\,\AA, 8542.09\,\AA, and 8662.14\,\AA, with $\sim 8542$\,\AA chosen as the central wavelength, hereafter denoted as \ion{Ca}{ii}\,NIR3) polarization. 

\subsection{High Continuum Polarization}
\label{s:cont}
The continuum polarization during the recombination phase of SNe~IIP has been shown to be relatively low, up to only a few tenths of a percent in the first 3 months after explosion \citep{leonard_is_2001,leonard_non-spherical_2006,chornock_large_2010}. 
However, there has been increasing evidence for diversity in photospheric-phase polarimetry of SNe~IIP (e.g., SNe\,2012aw, 2017ahn, and 2017gmr; \citealp{nagao_aspherical_2019,nagao_evidence_2021,dessart_polarization_2021}.)
In general, the low continuum polarization can be explained by a large opaque hydrogen envelope that  absorbs and reprocesses any light emanating from an asymmetric core. 
Following the photospheric phase, when their optical brightness falls off from the plateau \citep{eastman_theoretical_1994}, SNe~IIP have shown significant continuum polarization. \citet{leonard_non-spherical_2006} suggest that this is the result of the recombination front receding into the deeper He-rich layers of the ejecta revealing an asymmetric core. 

\citet{chugai_asymmetry_2006} proposes that the rapid rise in polarization at the beginning of the nebular phase originates from uneven ionisation by $^{56}$Ni plumes in the core. \citet{dessart_synthetic_2011} argue that this rise is linked to the ejecta becoming optically thin. This behaviour was explained in great detail for Type IIP SN\, 2004dj, 
which showed negligible continuum polarization ($p \leq 0.1$\%) until a sudden rise to $\sim 0.6$\% at $\sim 91$ days after the explosion \citep{leonard_non-spherical_2006}.
 
The rapid rise in polarization was coincident with the steepest decline of the $V$-band magnitude and has been suggested to be the result of an emerging asymmetric core. 
Thereafter, the continuum polarization of SN\,2004dj declined following a $t^{-2}$ law owing to the expansion of the ejecta and hence a decreasing optical depth. 

In this section, we discuss the temporal evolution of the continuum polarization measured during the photospheric phase of SN\,2021yja. 
We calculate the degree of linear polarization ($p$) and polarization angle (PA) as a function of wavelength for the first epoch of VLT FORS2 and five subsequent epochs of Lick/Kast spectropolarimetry for SN\,2021yja. 
We also estimate the continuum polarization for the first three months after the SN explosion. 
In this work, we compute the continuum polarization based on the $1\sigma$-error-weighted mean Stokes parameters over the spectral regions 6800--7200~\AA\ and 7820--8140~\AA. 
These wavelength ranges are selected based on the rationale presented by \citet{chornock_large_2010}. Both ranges are free of prominent and strongly polarized spectral features such as H$\alpha$, O~I $\lambda$7774, and \ion{Ca}{ii}\,NIR3. The choice also allows a direct comparison between SN\,2021yja and previous works \citep{chornock_large_2010, mauerhan_asphericity_2017, nagao_evidence_2021}. 
The uncertainty of the ISP was not included, considering its negligible contribution to the overall polarization (see Section~\ref{sec:isp}). 
The estimated levels of continuum polarization have also been corrected for bias arising from $p$ 
 being a positive-definite quantity following \citet{patat_error_2006} and \cite{maund_spectropolarimetry_2007}.

The continuum polarization and associated PA measured from the VLT epoch-1 observation on day 25 (Figs.~\ref{fig:qu_vlt_yi} and \ref{fig:pol_ep1})  are consistent with that of the Kast epoch-2 spectropolarimetry on day 30 (Fig. \ref{fig:pol_ep2}).  (Fig.~\ref{fig:pol_ep1} contains the same data as Fig.~\ref{fig:qu_vlt_yi}, but is shown in a the same way
as Figs. \ref{fig:pol_ep3}--\ref{fig:pol_ep6} in order to facilitate comparison of results from the different epochs.)
Such agreement is expected for a relatively short separation in time, and the measured continuum polarization is only mildly sensitive to spectral resolution.
The two sets of observations were carried out independently with a separation of five days with different instruments and under distinct observing conditions (e.g., airmass and seeing conditions, Table~\ref{tbl:specpol_log}). 
Both epochs exhibit a high level of polarization ($\gtrsim 0.8$\%) across the optical wavelength range and a strong depolarization near the emission centre of the H$\alpha$ profile. Similar features can also be identified in the Kast epoch-3 data at day 38 (Figure \ref{fig:pol_ep3}), which has a higher signal-to-noise ratio (S/N) compared to epoch 2. 
We present the spectropolarimetry of SN\,2021yja measured at days 57, 66, and 95 in Figures~\ref{fig:pol_ep4}, ~\ref{fig:pol_ep5}, and ~\ref{fig:pol_ep6}, respectively. 
The polarization spectra measured for all six epochs are also compared in Figure \ref{fig:all_epochs_pol_rows}. The temporal evolution of the continuum polarization between days $+$25 and $+95$ is shown in Figure~\ref{fig:cont}. 

The linear polarization originating from an aspherical density profile in the ejecta follows the relation $p \propto {\rm sin}^2\theta$, where $\theta$ is the viewing angle that measures the angular separation between the major axis of the ellipsoidal density profile and the line of sight. Therefore, the measured continuum polarization level estimates the lower limit of the degree of asphericity of the electron-scattering atmosphere.
Continuum polarization observed during the photospheric phase of SNe~II may arise from electron scattering in an aspherical photosphere or photons scattered by aspherically distributed dust in the CSM \citep{nagao_aspherical_2019}. The latter is expected to display a strong wavelength-dependent level of polarization --
in particular, enhanced polarization toward shorter wavelengths owing to a higher fraction of dust grains at relatively small sizes \citep{nagao_multi-band_2018}. 
 We fit a Serkowski's law to the polarization spectrum of SN\,2021yja at day $+$25, excluding spectral regions blueward of 5400\,\AA\ (which exhibit multiple polarized features) and H$\alpha$. The fittings failed to converge to a peak wavelength and did not give a better result compared to a horizontal line.
Therefore, we claim that no such polarization wavelength dependency can be identified in SN\,2021yja. 
We conclude that the continuum polarization of SN\,2021yja before the tail phase can be mostly attributed to the presence of an aspherical photosphere.

In our first epoch of VLT spectropolarimetry of SN\,2021yja at 25 days after the explosion, we measure a continuum polarization $p_{\text{cont}} = 0.87 \pm 0.06$\%. 
For Kast observations on days +30, +38, +57, +66, and +95, we measure $p_{\text{cont}} = 0.80 \pm 0.06$, $0.88 \pm 0.03$, $0.82 \pm 0.04$, $0.66 \pm 0.03$, and $0.55 \pm 0.06$\%, respectively. These results, along with the continuum polarization angle, are presented in Table \ref{tbl:contpol_log}. 
By approximating the density structure ($\rho$) of the SN\,2021yja ejecta with an oblate ellipsoidal electron-scattering-dominated atmosphere (e.g., $\rho \propto r^{-n}$, where $r$ is the distance from a given point to the mass centre), one could estimate its axis ratio for different density slopes $n$. 
The higher the value of $n$ for a given Thomson scattering depth $\tau$, the larger the asphericity required to achieve a certain level of polarization \citep{hoflich_asphericity_1991}, since multiple scattering may become dominant under the condition of a steep density gradient, which diminishes the geometric signature possessed by the photosphere. 

Assuming $n=2$ and an optical depth $\tau=1$ to represent the photosphere of SN\,2021yja a few weeks after the explosion, we estimate an asymmetry of $\sim 15$\% or an axis ratio of 1.15:1 when viewed equator-on (see Fig. 4 of \citealt{hoflich_asphericity_1991}). A higher degree of asphericity of $\sim 35$\% or an axis ratio of 1.35:1 can be estimated for $n=3$ and $\tau=5$ that represent a steeper density profile (see Fig.~4 of \citealp{yang_young_2020}).  

The continuum polarization is roughly constant at the 2$\sigma$ level until day 57. In the following epochs, we identify a decrease in the continuum polarization over time. Such a decreasing polarization level across the continuum during the plateau phase of a Type IIP SN is not only unprecedented (see Section \ref{s:intro}), 
but also contradictory to the consistently low continuum polarization before a rapid rise at the time of transition to the nebular phase. We suggest that the decrease in continuum polarization of SN\,2021yja during the photospheric phase can be a result of the photosphere receding into inner, more spherical layers, the optical thinning of the ejecta,  or a more pole-on alignment of the inner ellipsoid. 
Alternatively, the scenario that may explain this temporal evolution can be simplified as a two-component model.
The presence of the secondary component is hinted by the rotation of the polarization PA during the steadily decreasing phase of $p$ (see Section \ref{s:csm}).

As presented in Table~\ref{tbl:contpol_log} and Figure~\ref{fig:cont}, 
during the early plateau phase between days +25 and +66, we measure a time-invariant PA across the wavelengths that define the continuum. 
However, the continuum PA rotated to $124.0^\circ \pm 2.3^\circ$ on day 95, corresponding to $\sim 45$ days before the plateau dropoff. 
The interior layers of SN\,2021yja have an axis of symmetry that appears slightly rotated with respect to the outer H-rich layer. 
The high polarization measured over the optical wavelength range suggests a persisting aspherical electron-scattering envelope.
This object adds to an emerging heterogeneous sample of Type IIP polarimetry within the last decade, signaling a paradigm shift.

\subsection{Comparison with Other SNe}
We compare the unusually high early-time continuum polarization of 
SN\,2021yja with the well-studied Type IIP SNe
1999em (5050--5950\,\AA, \citealt{leonard_is_2001}), 2004dj (6800--8200\,\AA, \citealt{leonard_non-spherical_2006}), 2007aa (6800--7200\,\AA\ and 7820--8140\,\AA, \citealt{chornock_large_2010}), 2012aw (6900--7200\,\AA, \citealt{dessart_multiepoch_2021}, $R$-band Cousins filter [$\lambda_{R_{\text{eff}}}= 0.67 \mu$m], \citealt{kumar_broad-band_2014}), 2013ej (7800--8150\,\AA, \citealt{mauerhan_asphericity_2017}; 6800--7200\,\AA\ and 7820--8140\,\AA, \citealt{nagao_evidence_2021}), and 2017gmr (PMOS FILT\_815\_13 filter [$\lambda_0$ = 8150\,\AA, FWHM = 130\,\AA], \citealt{nagao_aspherical_2019}) in Figure \ref{fig:sne_comp}. 
Note that the different choices of wavelength ranges and filters used to determine the continuum polarization may 
introduce systematic uncertainties that are difficult to characterise. 
We also remark that the light curve of SN\,2013ej exhibits a faster decline compared with typical SNe~IIP, and its spectroscopic evolution more resembles the average behaviour of SNe~IIL \citep{valenti_diversity_2016}.
For this reason, we consider SN\,2013ej a transitional Type IIP/L SN throughout this paper.

As presented in Figure~\ref{fig:sne_comp}, except for SNe\,2013ej and 2021yja, the continuum polarization measured for other SNe~IIP is generally low during the period $t \approx 120$--40 days before the SN brightness falls from the plateau. 
During this phase, most SNe~IIP show an increase in the continuum polarization over time.
However, both SNe\,2013ej and 2021yja displayed a significant level of continuum polarization, $p_{\text{cont}} \approx 1$\% during the first few weeks after the explosion. 
We represent the phase in terms of days relative to the midpoint of the plateau falloff, as opposed to days from the explosion, to more effectively illustrate the progression of the photosphere into the ejecta.

In order to explain the high polarization observed for SN\,2013ej, \citet{mauerhan_asphericity_2017} suggest that there is interaction between the SN ejecta and an aspherical CSM as well as dust scattering, 
whereas \citet{nagao_evidence_2021} propose a combination of an aspherical explosion and a CSM-interaction component. 
We caution applying these interpretations to explain the similarly high polarization observed for SN\,2021yja given the lack of evidence for dense CSM from ultraviolet and optical spectroscopy \citep{hosseinzadeh_weak_2022,vasylyev_early-time_2022}. 
However, we do not rule out the existence of low-density CSM adjacent to the hydrogen envelope in SN\,2021yja given the interpretations made by
\citet{hosseinzadeh_weak_2022} and \citet{kozyreva_circumstellar_2022}; see Section~\ref{s:csm} for further discussion.

CCSNe have been associated with the production of relativistic jets that either puncture the outer layers or are choked by a dense hydrogen envelope \citep{khokhlov_jet-induced_1999, couch_aspherical_2009, papish_call_2015, piran_relativistic_2019}. 
Modeling of these jets has shown that they can induce a global asphericity in both the hydrogen envelope and the SN core 
\citep{hoflich_aspherical_2001, wang_axisymmetric_2002, maund_spectropolarimetry_2007}. For SNe~IIP, the bipolar jet model predicts the radioactive $^{56}$Ni to be confined to the inner regions, where it would ionise the surrounding material, producing an aspherical photosphere \citep{hoflich_star_odyssey, leonard_non-spherical_2006}. This, however, begs for an explanation of why many SNe~IIP and other noninteracting or weakly-interacting SNe (e.g., SN\,2004dj) display extremely low continuum polarization during their photospheric phases, unlike SN\,2021yja and SN\,2013ej. 

SN\,2017ahn gives another case that shows a rather unique temporal evolution of the continuum polarization when compared to SN\,2021yja and other SNe~IIP. \citet{nagao_aspherical_2019} observe a low continuum polarization during both the photospheric and early nebular phases, suggesting that the line of sight is close to the polar axis of the SN. 
However, we do not expect viewing-angle effects to be solely responsible for the observed diversity in the temporal evolution of the continuum polarization in SNe~IIP. 
For example, the polarization observed in SNe\,2012aw, 2013ej, and 2017gmr all display a continuous, monotonic rise before their optical light curves fall off the optical plateau, while SN\,2004dj maintains a continuum polarization of zero until the sudden rise at the end of the plateau.

One interpretation of the rising continuum polarization is that the ballistic jet is able to break far enough into the hydrogen envelope to produce an early rise in polarization during the photospheric phase
\citep{mauerhan_asphericity_2017, nagao_aspherical_2019, nagao_evidence_2021}. This scenario presents a counterexample to SN\,2004dj, where the ballistic jet would not punch far into the hydrogen envelope and fail to produce a steady rise in polarization before the onset of the nebular phase.  
We emphasise that the continuum polarization observed for SN\,2012aw and SN\,2017gmr rose to $>0.5\%$ much later in the photospheric phase than for SN\,2021yja, which already had a polarization of $\sim 0.8\%$ by day +30, similar to SN\,2013ej. 
Spectroscopic and polarimetric follow-up observations of SN\,2021yja during the nebular phase will be important for determining the origin of the global asymmetry, while peering into deeper layers of the core.

To summarise, SN\,2021yja is spectroscopically similar to other SNe~IIP that lack evidence of strong ejecta-CSM interaction, but are expected to have at least some pre-existing CSM produced by the RSG progenitor (see \citealt{morozova_measuring_2018, hillier_photometric_2019}). 
However, the observed high continuum polarization suggests the presence of an aspherical, extended hydrogen envelope with possible contributions from an additional heating source, such as interaction with a low-density CSM.
An ellispoidal geometry of the hydrogen envelope is compatible with the presence of bipolar relativistic jets.

\subsection{Additional Heating Source} 
\label{s:csm}
Following an approach similar to that of \citet{nagao_evidence_2021}, we discuss the possibility of an additional heating source (along with radioactive decay), namely 
CSM interaction, which can explain the high early-time continuum polarization and its steady decline over the photospheric phase. 


The main assumption is that the observed continuum polarization consists of two components: an aspherical explosion intrinsic to the SN and Thomson scattering in the asymmetric CSM. 
The latter is assumed to be constant over time -- the Stokes parameters that describe the CSM, 
$q_{\text{CSM}}$ and $u_{\text{CSM}}$, do not change appreciably. 
The first epoch, which shows the highest continuum polarization, is considered as being completely caused by electron scattering in the CSM component.
Choosing a constant CSM component may introduce a systematic error, especially for the later epochs. Therefore, we take caution in the interpretation of the results following such a procedure that separates the contribution from CSM scattering and an aspherical central energy source.
These assumptions are motivated by the observed pattern of weakly interacting SNe showing low intrinsic polarization at early times (see Section 3.3). Furthermore, \citet{nagao_evidence_2021} argue for the presence of an aspherical CSM in order to explain the high early-time continuum polarization of SN\,2013ej. The procedure is carried out by subtracting the assumed CSM Stokes parameters from the observed Stokes parameters that were used to calculate the continuum polarization in Section \ref{s:cont}. We estimate the uncertainty of the intrinsic polarization following the approximation $\sigma_{\text{CSM}} \approx 1.4 \,\sigma_{\text{obs}}$, where $\sigma_{\text{obs}}$ is the observed continuum polarization uncertainty. 

\begin{figure}
    \centering
    \includegraphics[width=0.53\textwidth]{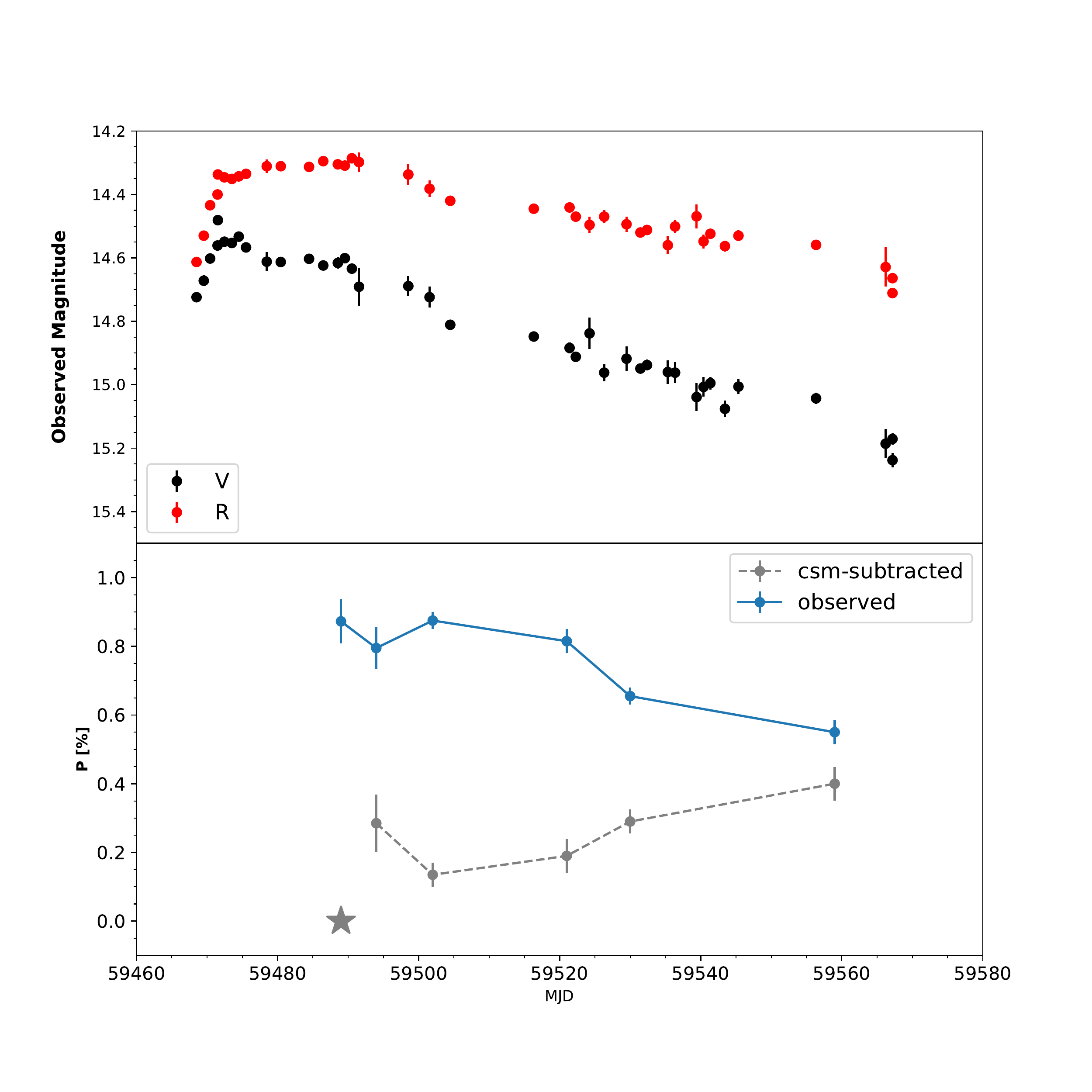}
    \caption{Upper panel is the same as in Figure \ref{fig:cont}. Temporal evolution of the CSP-corrected broad-band polarization of SN\,2021yja as compared to the observed values. 
    Error bars indicate 1$\sigma$ uncertainties, which do not account for the systematic uncertainty of the CSP correction. The grey star represents the assumed zero level of polarization on day 25.
    The presented values indicate the continuum level of the polarization calculated across the same wavelength range as in Figure~\ref{fig:cont}.}
    \label{fig:csm_sub}
\end{figure}

We present the temporal evolution of the continuum polarization of SN\,2021yja after subtracting the circumstellar polarization (CSP) component in Figure~ \ref{fig:csm_sub}. 
The thereby implied intrinsic polarization of the SN shows a monotonic increase from the assumed level of $p_{\text{cont}}=0\% $ at day 25 to $p_{\text{cont}} \sim 0.4\% $ at day 95. 
This temporal evolution is similar to that of SN\,2017gmr (see Fig.~\ref{fig:sne_comp}). 
In fact, at similar phases the CSP-corrected SN\,2021yja and observed SN\,2017gmr polarization were roughly the same. For example, at 45 days before the end of the plateau, both SNe have a continuum polarization value of $p_{\text{cont}} \approx 0.4\% $. 
\citet{nagao_aspherical_2019} suggest that this early rise in continuum polarization (before the onset of the nebular phase) can be explained by asymmetries not only in the helium core, but also in the hydrogen envelope. 

We extend this argument to explain the temporal evolution of the intrinsic polarization of SN\,2021yja. This smooth transition between the outer and inner geometric configurations is compatible with a jet-driven explosion, 
where the relativistic jet breaks out of the helium core into the hydrogen envelope \citep{wang_bipolar_2001, couch_aspherical_2009}. 
The temporal evolution of SN\,2012aw is delayed relative to both the observed polarization of SN\,2017gmr and the CSP-corrected polarization of SN\,2021yja. 
Additional polarimetry of SN\,2021yja during the plateau dropoff and the nebular phase is crucial for determining the geometry of the helium core (Nagao et al. 2023, in prep.). 

\subsection{Line Polarization and the Stokes $q-u$ Diagrams}
\label{s:qu_diagrams}
\begin{figure*}
\begin{subfigure}{\textwidth}
  \centering
  \includegraphics[width=\linewidth]{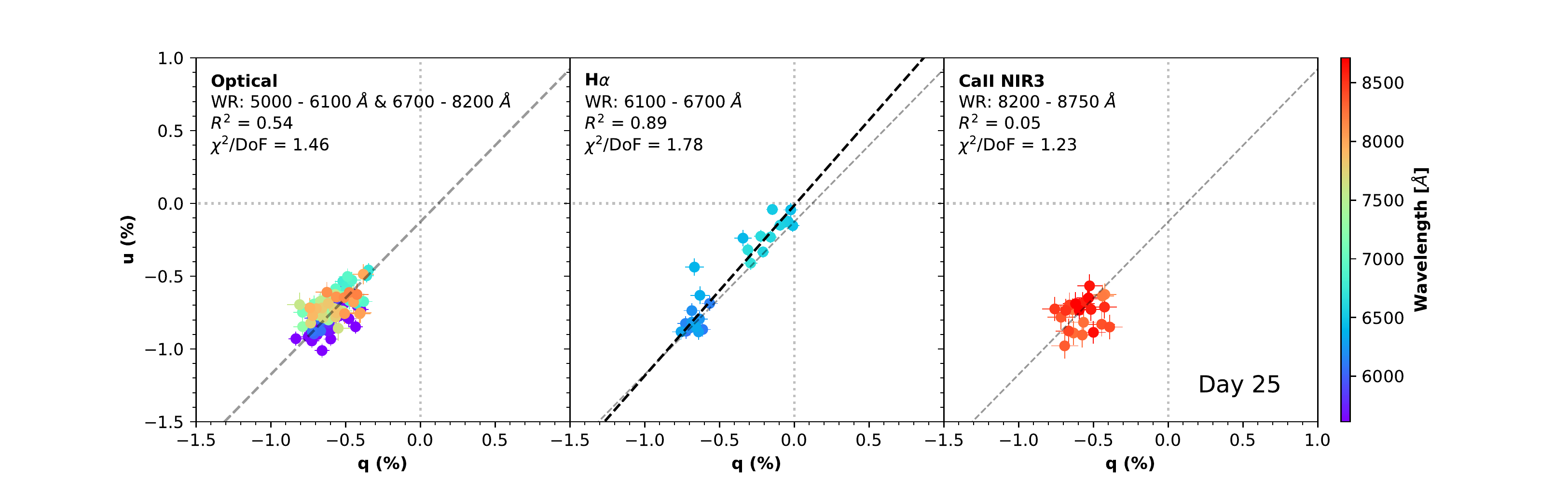}
  \label{fig:sfig1}
\end{subfigure}
\begin{subfigure}{\textwidth}
  \centering
  \includegraphics[width=\linewidth]{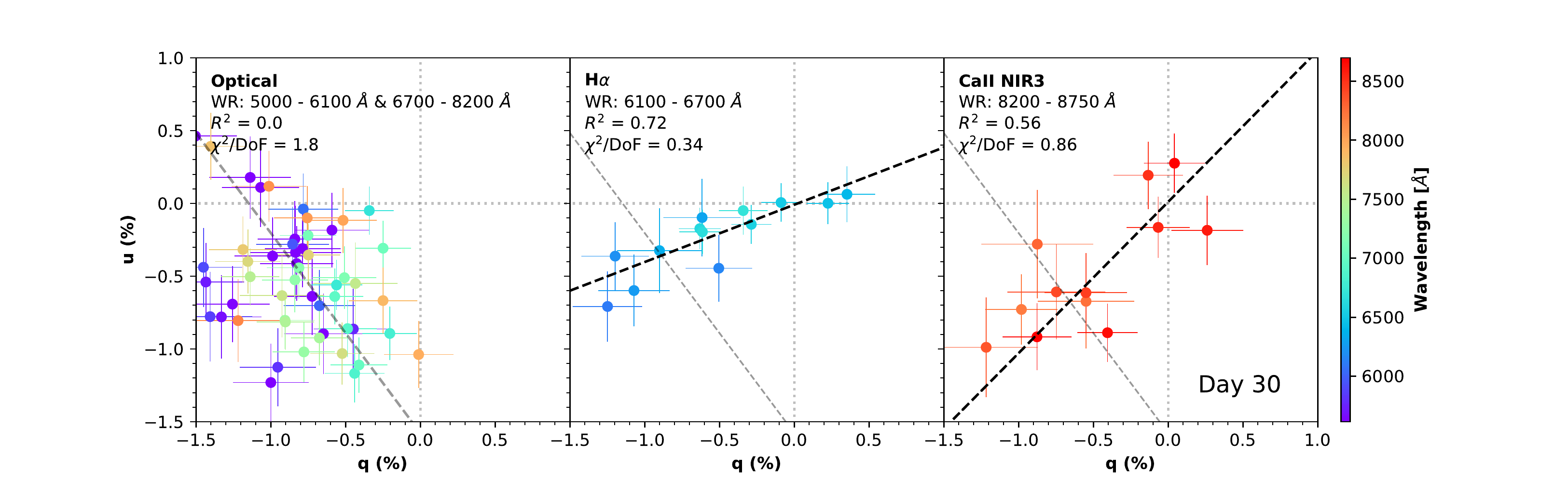}
  \label{fig:sfig2}
\end{subfigure}
\begin{subfigure}{\textwidth}
  \centering
  \includegraphics[width=\linewidth]{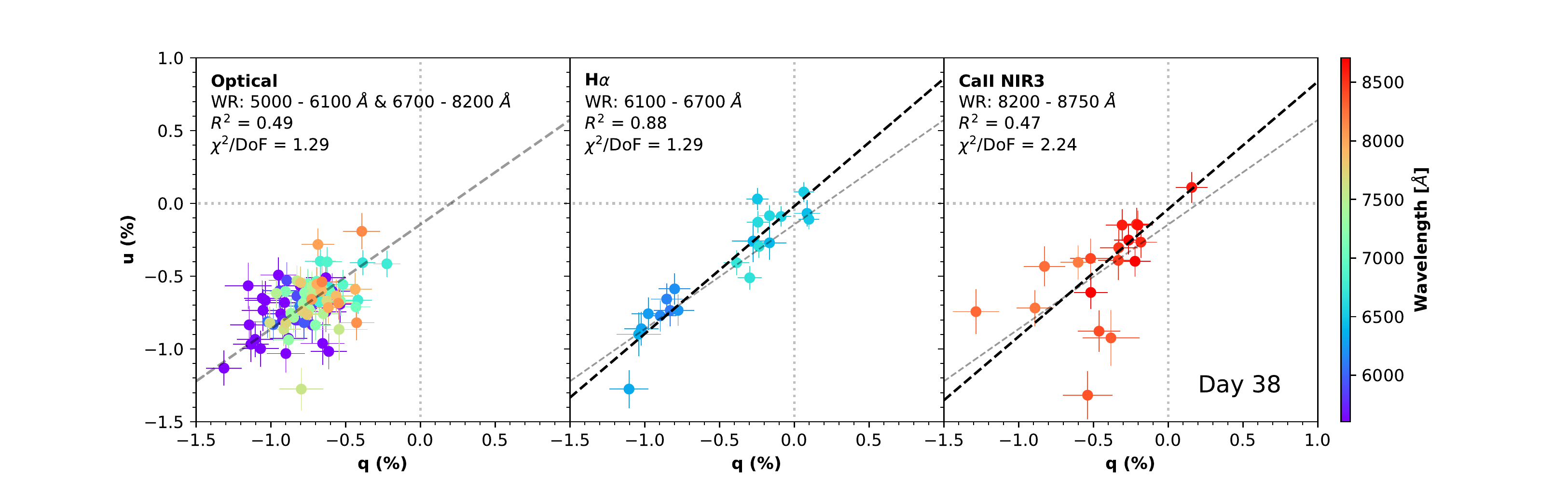}
  \label{fig:sfig3}
\end{subfigure}
\caption{Stokes parameters of SN\,2021yja displayed on the $q-u$ plane. The polarization data have been rebinned to 25\,\AA, 50\,\AA, and 30\,\AA\ for the observations at epoch 1 (day 25), epoch 2 (day 30), and epoch 3 (day 38), respectively. 
Error bars represent 1$\sigma$ uncertainties. Panels from left to right show the optical (5000--6100\,\AA\ and 6700--8200\,\AA), H$\alpha$ (6100--6700\,\AA), and \ion{Ca}{II}\,NIR3 (8200--8750\,\AA) regions. 
The dashed grey line in the second and third panels indicates the optical dominant axis computed from data in the ranges 5000--6100\,\AA\ and 6700--8200\,\AA. 
Bold, dashed lines in the second and third panels fit the dominant axis for the H$\alpha$ and the \ion{Ca}{II}\,NIR3 profiles, corresponding to a velocity range from roughly $-$21,200 to $+$6200\,km\,s$^{-1}$ and $-$12,000 to $+$7300\,km\,s$^{-1}$, respectively.
}
\label{fig:qu1}
\end{figure*}

\begin{figure*}
\begin{subfigure}{\textwidth}
  \centering
  \includegraphics[width=\linewidth]{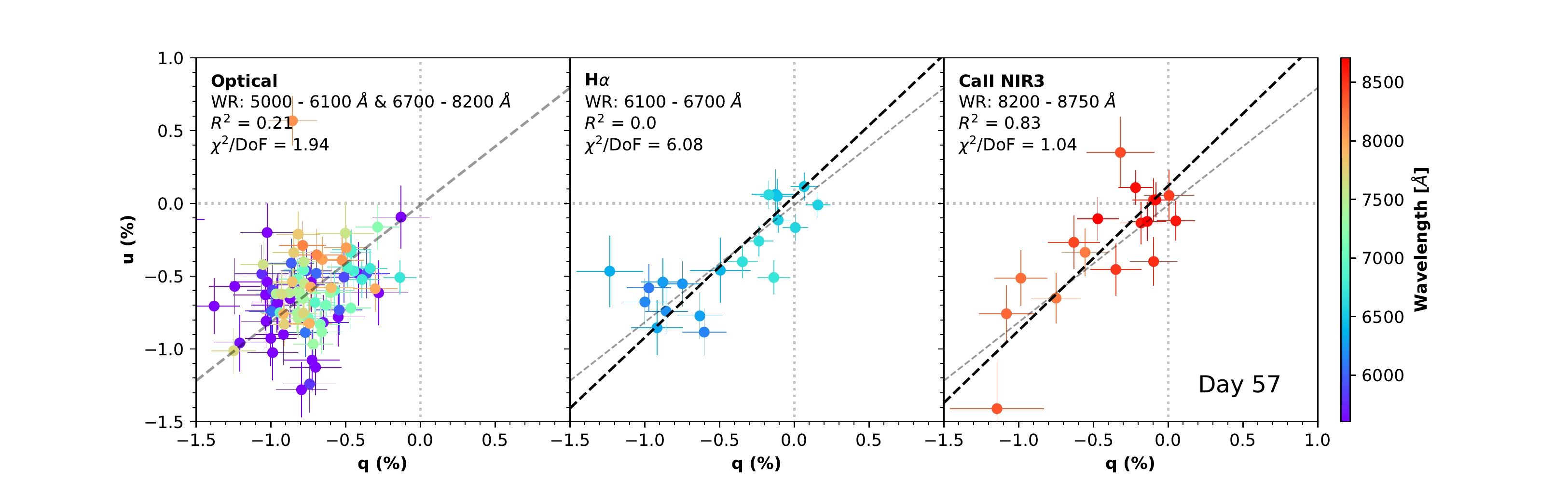}
  \label{fig:sfig1}
\end{subfigure}
\begin{subfigure}{\textwidth}
  \centering
  \includegraphics[width=\linewidth]{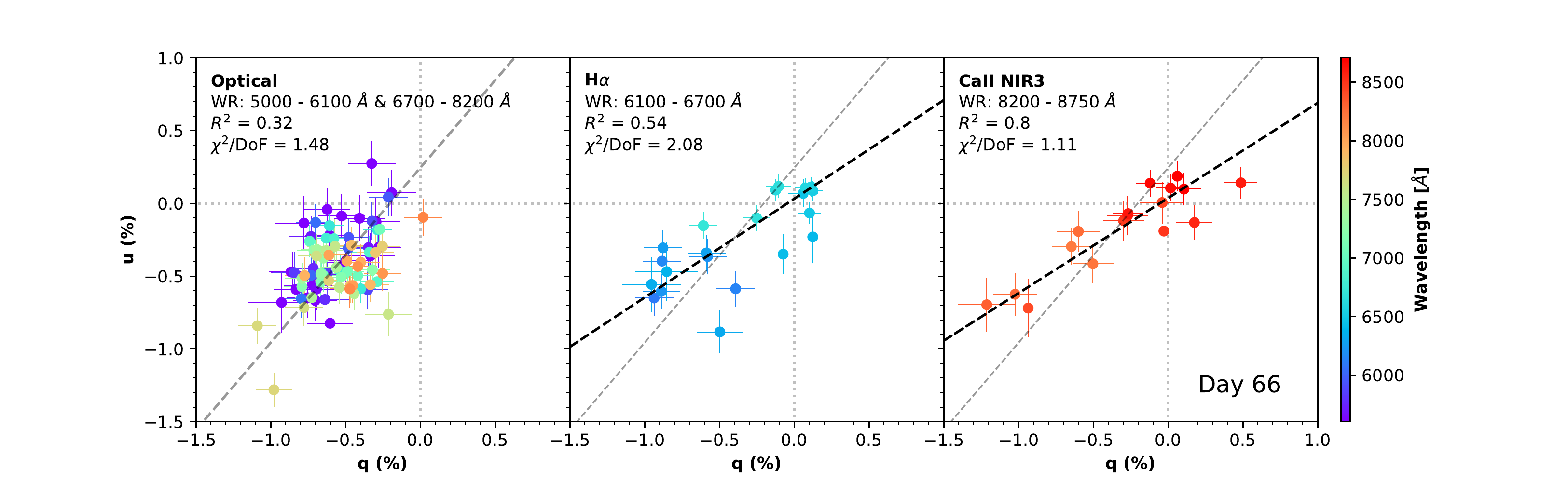}
  \label{fig:sfig2}
\end{subfigure}
\begin{subfigure}{\textwidth}
  \centering
  \includegraphics[width=\linewidth]{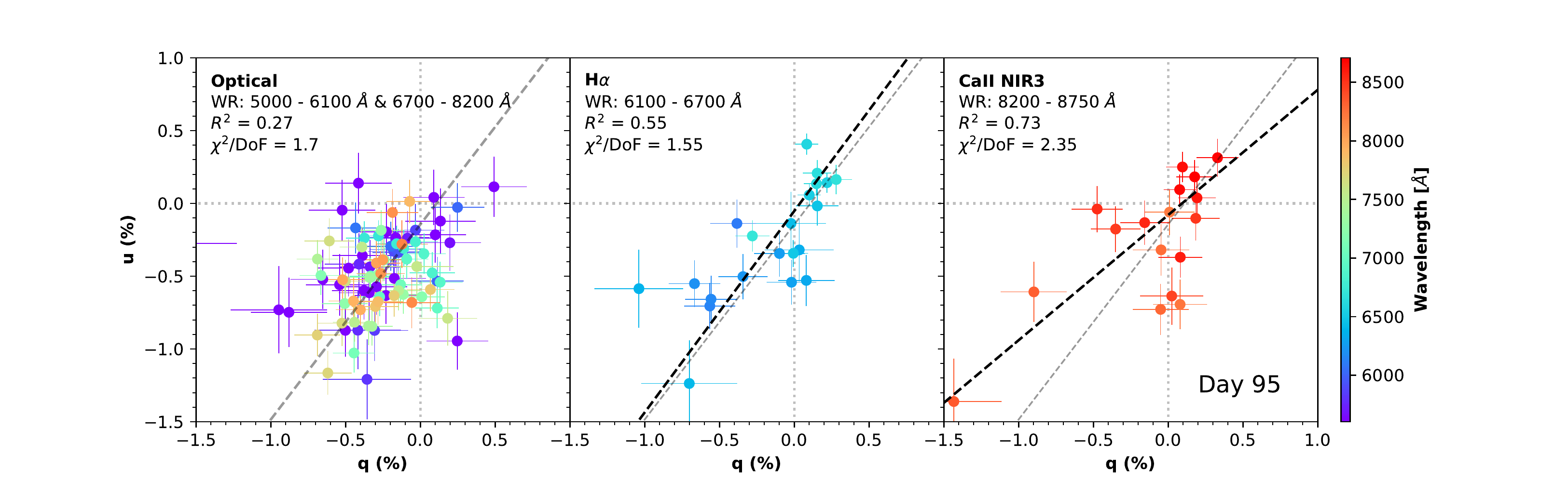}
  \label{fig:sfig3}
\end{subfigure}
\caption{Similar to Figure~\ref{fig:qu1} but for epochs 4 (day 57, top row), 5 (day 66, middle row), and 6 (day 95, bottom row). Data are binned to 30\,\AA.} 
\label{fig:qu2}
\end{figure*}


Polarization across prominent spectral lines can be used to map the distribution of individual elements within the SN ejecta. 
In order to better illustrate the geometry of various elements and compare their symmetry axes with that of the SN ejecta, we adopt a mathematically convenient alternative expression to $p$ and PA by presenting the fractional Stokes $q$ and $u$ parameters per wavelength bin on a two-dimensional plane \citep{wang_bipolar_2001}. 
When the ejecta share a common axis of symmetry, 
their values for the $q$ and $u$ Stokes parameters fall on a straight line in the $q-u$ plane, colloquially known as the dominant axis \citep{wang_spectropolarimetry_2003,maund_spectropolarimetric_2010}. 

On the other hand, when ejecta depart from an axial symmetry, the values for $q$ and $u$ deviate from the dominant axis in the form of loops or a continuous rotation of the PA as a function of wavelength across the corresponding line profile. 
Loop structures in the $q-u$ plane are treated as indicators of ``clumpiness'' in the ejecta and are perhaps most well studied for the \ion{Ca}{II} NIR3 
line \citep{kasen_analysis_2003}. 
Multidimensional hydrodynamics simulations of the SN atmosphere using Monte Carlo radiative-transfer methods will be essential to quantify the size, distribution, and optical depth of these clumps.

A commonly observed behaviour across prominent Balmer lines is the progressive depolarization toward the centre of the emission features. 
The emission process scrambles preferentially oriented electric-field vectors of the incoming (absorbed) photons \citep{jeffery_analysis_1991}. 
As the SN ejecta expand over time, the column density drops, and the scattering opacity decreases. In the case of the optically thick regime ($\tau \textgreater 1$), the decline of multiple scattering will initially result in an increase of $p_{\text{cont}}$ until the optical depth of electron scattering drops to $\sim 1$; then $p_{\text{cont}}$ will begin to decrease monotonically (see Fig.~1 of \citet{hoflich_asphericity_1991}).
Apart from the depolarized Balmer features, we also note that major absorption lines, such as \ion{Ca}{ii}\,NIR3, exhibit their highest polarization close to the absorption minimum of the feature \citep{mccall_are_1984, leonard_is_2001}. 
The line-forming region sits just above the photosphere, thus probing a different region of the ejecta.

The left column of Figures~\ref{fig:qu1}--\ref{fig:qu2} shows the spectropolarimetry of SN\,2021yja presented on the $q-u$ plane for all six epochs from days $+$25 to $+$95. Black dashed lines fit the dominant axis to the observed polarization of SN\,2021yja in the wavelength range $4600 \leq \lambda \leq 8900$\,\AA, 
representing the direction of axial symmetry. Data points within the wavelength covered by the significantly depolarized Balmer lines, the highly polarized \ion{Ca}{II}\,NIR3, and the major telluric bands were excluded.

The middle and the right columns of Figures~\ref{fig:qu1}--\ref{fig:qu2} are the same as the left column but for the H$\alpha$ line from $\sim -21,200$ to +6200\,km\,s$^{-1}$ with respect to a central wavelength of $\lambda_{0}^{\rm H\alpha}=6564.6$\,\AA\ and the \ion{Ca}{ii}\,NIR3 profile from $\approx -12,000$ to $+$7300\,km\,s$^{-1}$ with respect to a central wavelength of $\lambda_{0}^{\rm CaII}=8542$\,\AA, corresponding to the middle component of the triplet. Linear fits to the displayed data points that cover the H$\alpha$ and the \ion{Ca}{II}\,NIR3 profiles are represented by the grey-dotted lines in the middle and the right columns, respectively. 

The H$\alpha$ and \ion{Ca}{II}\,NIR3 lines show similar axial symmetry with the optical continuum, which was chosen over the range 4600--8900\,\AA\, (ignoring major spectral lines). 
At all epochs, the PA across the H$\alpha$ profile also exhibits little wavelength dependence.
Moreover, the PA of the H$\alpha$ is overall persistently aligned with the dominant axes fitted over the optical wavelength range (4600--8900\,\AA, ignoring H$\alpha$ and \ion{Ca}{II}\,NIR3). 
Therefore, we suggest that the continuum-emitting region in the ejecta and the H-rich envelope share a similar axial symmetry. 

Nearly complete depolarization occurs near the \ion{Ca}{II}\,NIR3 emission peak on day 30, whereas on day 25, the polarization across the line follows the continuum. 
The lack of a strong \ion{Ca}{ii} line feature in both the flux and polarization spectra for the earlier VLT epoch 
suggests that the receding photosphere has not yet reached the calcium layer. Strong line polarization in H$\alpha$ and 
\ion{Ca}{ii}\,NIR3 does not appear until day +38. 

\begin{figure}
    \includegraphics[width=0.48\textwidth]{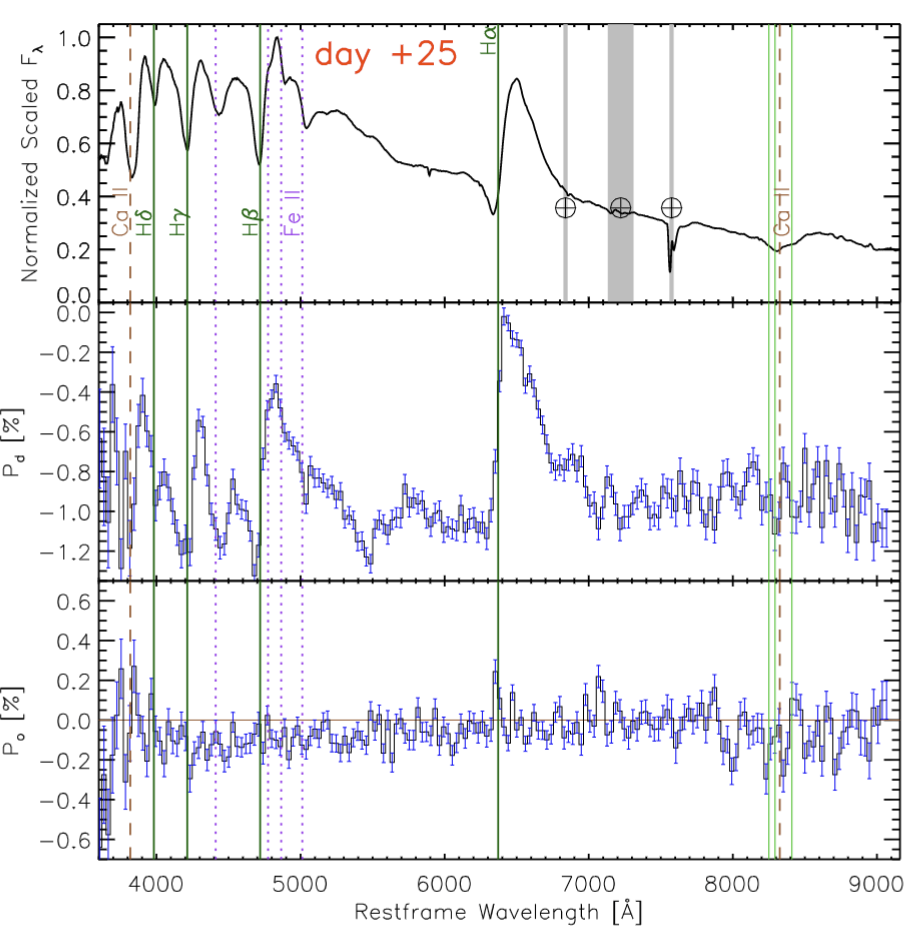}
    \caption{Normalised total-flux spectrum (upper panel) together with the principal-component analysis of SN\,2021yja spectropolarimetry at day +25. The polarization spectra projected onto the dominant and the orthogonal axes are shown in the middle and the bottom panels, respectively. Some major telluric lines are labeled by $\earth$ and vertical grey-shaded bands.}
    \label{fig:pca}
\end{figure}

This can also be illustrated by projecting the polarization onto the dominant axis ($P_{\rm{d}}$) that characterises the direction of global asphericity (see the middle panel of Fig.~\ref{fig:pca}). The residual will be the polarization signal orthogonal to the dominant axis ($P_{\rm{o}}$), which represents the deviations from axial symmetry (see the bottom panel of Fig.~\ref{fig:pca}). 
At day +25, all major polarization modulations are observed in the dominant polarization component, while the signals in the orthogonal polarization component do not show significant deviation from zero. Such a low residual across the entire observed wavelength range implies a well-defined symmetry axis shared by the continuum and various line-forming regions \citep{wang_spectropolarimetry_2008}.

On day +95, we observe a slight rotation in both the continuum dominant axis and H$\alpha$ (Fig.~\ref{fig:qu2}). However, the rotation is small and may be caused by the emergence of other spectral lines along the chosen continuum wavelength range. 
Therefore, we conclude that the dominant axes of both H$\alpha$ and \ion{Ca}{II}\,NIR3 share a similar direction with that of the SN ejecta. The common symmetry axis of SN\,2021yja does not rotate significantly during the photospheric phase, suggesting a global asphericity in the hydrogen-rich envelope. Additionally, we observe a loop structure in the $q-u$ plot for \ion{Ca}{II}\,NIR3  on days 38 and 95, which may suggest a clumpy distribution of calcium deeper in the ejecta. The low S/N complicates the determination of loop structures for the H$\alpha$ line.  Additional spectropolarimetry during the nebular phase will be necessary to determine if a greater deviation from spherical symmetry exists in the inner He-rich core, which may reflect an enhanced instability during the SN explosion (Nagao et al. 2023, in prep.).


SN\,2021yja shows an increase in the level of polarization across the \ion{Ca}{II}\,NIR3 line throughout the six epochs of spectropolarimetry. For example, the peak polarization across \ion{Ca}{II}\,NIR3 grows almost monotonically from $\sim 1.5$\% on day $+$38 to $\gtrsim 3$\% on day +95. 
Meanwhile, the continuum polarization decreases over time, indicating a considerable decline in electron-scattering opacity in the SN ejecta. 
Therefore, given opacity distributions that do not change dramatically across the radial direction, the corresponding line polarization is expected to drop, contradicting the observed polarization behaviour of SN\,2021yja (Figs.~\ref{fig:pol_ep1}--\ref{fig:pol_ep6}). 
A similar trend of increase of \ion{Ca}{II}\,NIR3 polarization in the nebular phase has been reported for the Type Ia SNe\,2006X \citep{patat_vlt_2009} and 2021rhu \citep{yang_spectropolarimetry_2022}. 
This result is unexpected under the framework that line polarization is produced primarily when the element opacity unevenly masks the underlying photosphere. 
Additionally, as the ejecta expand over time, it becomes progressively unlikely for a thermally-emitted photon to be destroyed by absorption. In other words, a high and continuously rising line polarization is also unexpected given the reduced thermalisation depth over time. 
Therefore, apart from the line polarization created through an interplay between line occultation and thermalisation \citep{Hoeflich_Yang_2023}, other mechanisms would be essential to explain the late-time rise of the \ion{Ca}{ii}\,NIR3 polarization.

An alternative explanation is that the late-time increase of the \ion{Ca}{II}\,NIR3 line polarization was the result of the geometric alignment of the ground and metastable states of this ion through photoexcitation by an anisotropic radiation field \citep{yang_spectropolarimetry_2022}. 
The angular momentum of atoms can be realigned through magnetic precession in a weak magnetic field under the condition that the Larmor precession rate is greater than the photoexcitation rate \citep{happer_optical_1972,landolfi_resonance_1986}. 
This type of magnetic alignment leads to an uneven distribution of atoms across magnetic sublevels and produces polarized radiation \citep{yan_polarization_2006, yan_tracing_2012}. 
Therefore, the presence of the polarized lines whose corresponding atomic states can be magnetically aligned may provide an opportunity to infer the magnitude and configuration of the magnetic field. 
For example, all three of the \ion{Ca}{ii}\,NIR3 triplet components, with air wavelengths of
8498.02\,\AA\ ($2D_{3/2} \rightarrow 2P_{3/2}$), 8542.09\,\AA\ ($2D_{5/2} \rightarrow 2P_{3/2}$), and 8662.14\,\AA\ ($2D_{3/2} \rightarrow 2P_{1/2}$), are metastable states. Each of them can be geometrically
aligned and individually polarized through photoexcitation when placed in an anisotropic radiation field. 

In the case of SN\,2021yja, the increase of \ion{Ca}{II}\,NIR3 polarization 
(from $\sim 1.5$\% on day $+$38 to $\gtrsim 3$\% on day +95) is not as significant as in the Type Ia SN\,2021rhu (from $\sim 0.4$\% around the peak luminosity to $\sim$2.5\% after 79 days, \citealp{yang_spectropolarimetry_2022}).
Additionally, the calcium opacity and/or the asphericity of its distribution may vary as the photosphere tomographically samples the inner layers of the ejecta. 
From day $+$38, the dominant axis over the \ion{Ca}{ii}\,NIR3 feature maintains good agreement with that of the continuum. A similar agreement is observed for the Type Ia SN\,2021rhu at 79 days after its $B$-band maximum (see Fig.~9 of \citealp{yang_spectropolarimetry_2022}). 
However, the dominant axis of \ion{Ca}{ii}\,NIR3 in SN\,2021rhu exhibits a rotation since 
its peak luminosity. 
The phases that cover such a rotation (i.e., $\sim 1$ week before the peak, around the peak, and a few weeks after the peak) correspond to when the photosphere recedes into the layers of complete oxygen burning, incomplete silicon burning, and the central Ni-rich core under nuclear statistical equilibrium, respectively \citep{Hoeflich_Yang_2023}. 
The rotation of the \ion{Ca}{ii}\,NIR3 dominant axis at early phases, therefore, may be accounted for by the distribution of \ion{Ca}{ii} opacity in the outer ejecta of the thermonuclear SN\,2021rhu. 
Despite the small sample size, the agreement between the dominant axes of the continuum and \ion{Ca}{ii} NIR3 in the two observed cases warrants further attention in future measurements. 
However, without detailed theoretical modeling, it is difficult to discern whether occlusion of the photosphere, optical pumping effects, or a combination of both explains the late-time rise of the \ion{Ca}{II}\,NIR3 polarization. 

\section{Conclusions}~\label{sec:conclusions}
Spectropolarimetric observations of SN\,2021yja during the photospheric phase reveal significant departures from global sphericity, suggesting a highly aspherical distribution of free electrons in the outer layers of the ejecta. 

The early continuum polarization was measured to be $p \approx 0.9\%$, placing SN\,2021yja among the most polarized known SNe~IIP during the photospheric phase. The observed continuum polarization shows a slight decline over time, which is unprecedented in the literature for SNe~II. We find almost complete depolarization near the centre of the H$\alpha$ emission line, suggesting a negligible amount of ISP contributed by the line-of-sight dust. Therefore, the unusually high observed continuum polarization is unlikely to be contaminated by the ISP. 
Meanwhile, both the orientation of the dominant axis and the continuum PA measured for SN\,2021yja remained constant from days 25 to 66, indicating a persistent axial symmetry from the outer and inner regions of the SN ejecta.

Following an approach similar to that of \citet{nagao_evidence_2021}, we assumed that the continuum polarization of SN\,2021yja during the early photospheric phase is dominated by ejecta interacting with the aspherical CSM. 
We then subtract this interaction component from the polarization measured at later phases to measure the continuum polarization instrinsic to the SN, which shows a steady rise throughout the photospheric phase. 

By modeling ultraviolet spectra of SN\,2021yja obtained between $+$9 and $+$21 days after the explosion, \citet{vasylyev_early-time_2022} find no significant flux contribution from the CSM. 
\citet{hosseinzadeh_weak_2022} do not find evidence for dense CSM from modeling the early-time light curves of SN\,2021yja using the model presented by \citet{sapir_uvoptical_2017}, 
suggesting a red supergiant progenitor with weak mass loss. 
However, the light-curve modeling by \citet{kozyreva_circumstellar_2022} as well as observations of X-rays, radio emission, a blue excess, and a rapid rise time suggest that at least some CSM interaction is occurring. 
These seemingly incompatible observations paint a confusing picture. We propose a compromise similar to that of \citet{hosseinzadeh_weak_2022}, where the density of the CSM is low enough to not manifest in spectral lines, but is sufficiently high to produce the observed properties listed above.

The spectropolarimetric observations could also be explained by an aspherical, extended hydrogen envelope, with a possible contribution by an additional heating source, such as interaction with a low-density CSM. 
However, the observed polarization can instead be interpreted as the result of an aspherical explosion with an extended hydrogen envelope, indistinguishable from low-density CSM adjacent to the SN. Additionally, the observed preferential axis of symmetry shared among the continuum and prominent spectral lines (H$\alpha$, \ion{Ca}{II}\,NIR3) is compatible with the presence of bipolar relativistic jets. 

The observed high polarization as early as 25 days after the explosion, the constant symmetry axis of the ejecta, and the lack of scatter in the $q-u$ diagram may suggest that a binary process distorted the massive hydrogen stellar envelope that was initially unperturbed by the explosion. In this scenario, Rayleigh-Taylor instabilities would not have reached the hydrogen envelope, where the photosphere is located at these early times. This interpretation adds to the ambiguity of the physical mechanisms responsible for the observed polarization. However, the consistency of this argument requires verification through nebular polarimetric observations and detailed multidimensional hydrodynamic simulations. Future work adding to the sample of SNe~IIP will be crucial in resolving the degeneracy in explosion models and geometries. A larger sample will also allow us to test correlations between the level of polarization and other SN parameters such as synthesised $^{56}$Ni mass and explosion energy (see discussion by \citealt{chornock_large_2010}). 

\section*{Acknowledgements} 
We thank Takashi Nagao and Ryan Chornock for fruitful discussions. 
We are grateful to the European Organisation for Astronomical Research 
in the Southern Hemisphere (ESO) for the generous allocation of observing time. The polarimetry studies in this work are based in part on observations made with the VLT at ESO's La Silla Paranal Observatory under program ID 105.20AU.002 and 108.228K.001. We especially thank the staff at Paranal for their proficient and highly motivated support of this project in service mode.
A major upgrade of the Kast spectrograph on the Shane 3\,m telescope at Lick Observatory, led by Brad Holden, was made possible through gifts from the Heising-Simons Foundation, William and Marina Kast, and the University of California Observatories.
KAIT and its ongoing operation were made possible by donations from Sun Microsystems, Inc., 
the Hewlett-Packard Company, AutoScope Corporation, Lick Observatory, the U.S. National 
Science Foundation, the University of California, the Sylvia \& Jim Katzman Foundation, and           
the TABASGO Foundation. We appreciate the expert assistance of the staff at Lick Observatory.  
Research at Lick Observatory is partially supported by a generous gift from Google. 

Generous financial support was provided to A.V.F.'s supernova group at U.C. Berkeley by Steven Nelson, Landon Noll, Sunil Nagaraj, Sandy Otellini, Gary and Cynthia Bengier, Clark and Sharon Winslow, Sanford Robertson, Alan Eustace, Frank and Kathleen Wood, the Christopher R. Redlich Fund, and the Miller Institute for Basic Research in Science (in which A.V.F. was a Miller Senior Fellow).

\section*{Data Availability}
The VLT spectropolarimetry data used in this work can be retrieved from the ESO Science Archive, program ID: 105.20AU.002 (PI Y. Yang). 
The reduced Lick data used in this work may be shared upon request to Sergiy S. Vasylyev (sergiy\_vasylyev@berkeley.edu) or Yi Yang (yi.yang@berkeley.edu).


\bibliographystyle{mnras}
\bibliography{2021yja_specpol} 




\appendix

\section{A test of the FORS2 spectropolarimetry precision}
~\label{section_sanity}
\begin{figure*}
    \centering
    \includegraphics[width=1.0\textwidth]{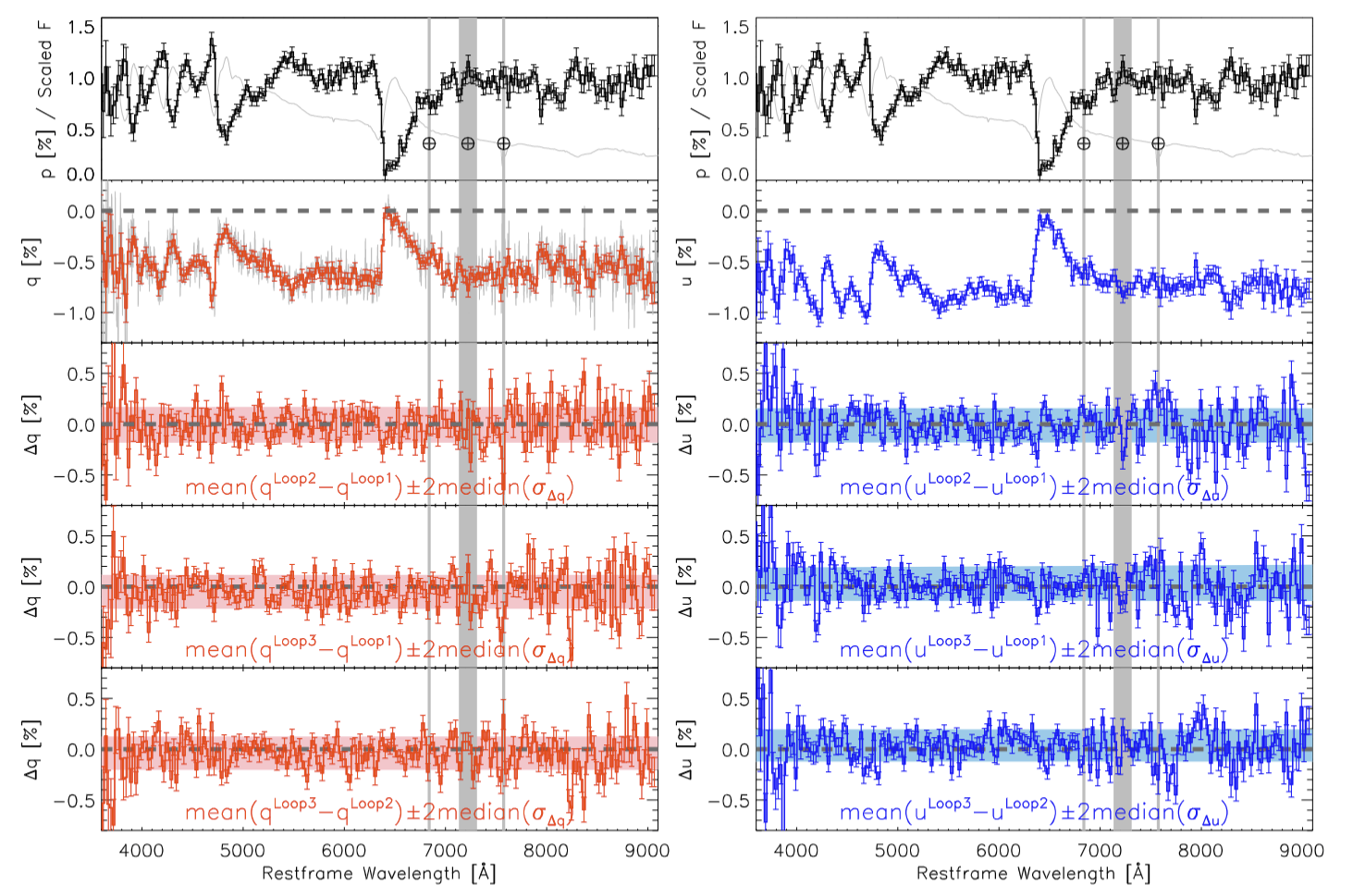}
\caption{
{\it Left column:} The five panels (from top to bottom) give (1) the polarization spectrum ($p$, black histogram) and the scaled-flux spectrum (grey line); (2) the normalised Stokes parameter $q$, (3--5), the difference between $q$ derived for loop\,2 and loop\,1, loop\,3 and loop\,1, and loop\,3 and loop\,2, respectively. The horizontal shaded bands represent the mean differences of $q \pm 2$ times the median uncertainty among all 25\,\AA\ bins presented in the figure across the observed wavelength range. {\it Right column:} similar to the left column but for $u$. Some major telluric lines are labeled by red-shaded lines marked with $\earth$.
~\label{fig:3loops}}
\end{figure*}

The first epoch of VLT spectropolarimetry was obtained by calculating the mean flux spectra of the o-rays and e-rays obtained at three consecutive loops of a set of four exposures at different retarder plate angles during the night of 2 Jan. 2021, with starting times of approximately 04:44, 05:08, and 05:50, respectively\footnote{The first and the second loops were obtained through ESO program 105.20AU.002, while the third loop was acquired through ESO program 108.228K.001.}. 
For the purposes of inspecting the precision and stability of FORS2 spectropolarimetry, we compare the $q$ and $u$ spectra for all three loops by presenting their differences in Figure~\ref{fig:3loops}. The pink and the light-blue shaded horizontal bands are vertically centred at the average $q$ and $u$ values across the entire observed wavelength range, respectively. 
The width of the shaded regions corresponds to twice the median uncertainty of the difference between the 25\,\AA\ binned Stokes parameters from the loops considered.

As shown in the third to the fifth rows of Figure~\ref{fig:3loops}, most of the bins in the difference spectra of the Stokes parameters computed from individual loops of spectropolarimetry are within two times the typical size of the statistical uncertainty. Therefore, we suggest that the Stokes parameters derived for the three consecutive loops agree within their 2$\sigma$ uncertainties. The outcome of such a sanity check also demonstrates the high precision and reproducibility of FORS2 spectropolarimetry, despite moderate variations in the airmass and the seeing conditions during the observation (see Table~\ref{tbl:specpol_log}). 


\label{lastpage}
\end{document}